\documentclass[journal]{IEEEtran}
\usepackage{indentfirst}
\usepackage[dvips]{graphicx}
\usepackage{amsfonts}
\usepackage{multirow}
\usepackage{amsmath,amsthm,amssymb}
\usepackage{color,changepage}
\usepackage{CJK}
\usepackage{algorithm}
\usepackage{algorithmic}
\usepackage{bbm}
\usepackage{verbatim}
\usepackage{makecell}
\usepackage{subfigure}
\usepackage{mathrsfs}
\usepackage{arydshln}
\usepackage{cases}
\usepackage{extarrows}
\usepackage{array}
\usepackage{bm}
\usepackage{pifont}
\usepackage[T1]{fontenc}
\usepackage{threeparttable}
\usepackage{booktabs}
\usepackage{mdwlist}
\allowdisplaybreaks[4]

\definecolor{newcolor}{rgb}{0.5,0,1}

\newcommand{\cmarkk}{\ding{56}}

\newcolumntype{I}{!{\vrule width 2pt}}

\newtheorem{Theorem}{Theorem}

\newtheorem{Lemma}{Lemma}

\theoremstyle{remark}
\newtheorem{Remark}{Remark}

\DeclareMathAlphabet{\mathpzc}{OT1}{pzc}{m}{it}

\definecolor{newcolor}{rgb}{0.5,0,1}

\begin{document}
\title{Multi-User Blind Symmetric Private Information Retrieval from Coded Servers}
\author{Jinbao Zhu, Qifa Yan, \IEEEmembership{Member, IEEE} and Xiaohu Tang, \IEEEmembership{Senior Member, IEEE}
\thanks{The authors are with the Information Security and National Computing Grid Laboratory, Southwest Jiaotong University, Chengdu 611756, China (email: jinbaozhu@my.swjtu.edu.cn, qifayan@swjtu.edu.cn, xhutang@swjtu.edu.cn).}
}
\maketitle

\pagestyle{empty}
\thispagestyle{empty}

\begin{abstract}
The problem of Multi-user Blind $X$-secure $T$-colluding Symmetric Private Information Retrieval from Maximum Distance Separable (MDS) coded storage system with $B$ Byzantine and $U$ unresponsive servers (U-B-MDS-MB-XTSPIR) is studied in this paper.
Specifically, a database consisting of multiple files, each labeled by $M$ indices, is stored at the distributed system with $N$ servers according to $(N,K+X)$ MDS codes over $\mathbb{F}_q$ such that any group of up to $X$ colluding servers learn nothing about the data files.
There are $M$ users, in which each user $m,m=1,\ldots,M$ privately selects an index $\theta_m$ and wishes to jointly retrieve the file specified by the $M$ users' indices $(\theta_1,\ldots,\theta_M)$ from the storage system, while keeping its index $\theta_m$ private from any $T_m$ colluding servers, where there exists $B$ Byzantine servers that can send arbitrary responses maliciously to confuse the users retrieving the desired file and $U$ unresponsive servers that will not respond any message at all.
In addition, each user must not learn information about the other users' indices and the database more than the desired file.
An U-B-MDS-MB-XTSPIR scheme is constructed based on Lagrange encoding. The scheme achieves a retrieval rate of $1-\frac{K+X+T_1+\ldots+T_M+2B-1}{N-U}$ with secrecy rate $\frac{K+X+T_1+\ldots+T_M-1}{ N-(K+X+T_1+\ldots+T_M+2B+U-1)}$ on the finite field of size $q\geq N+\max\{K, N-(K+X+T_1+\ldots+T_M+2B+U-1)\}$ for any number of files.
\end{abstract}

\begin{IEEEkeywords}
Private information retrieval, security, user privacy, blind privacy, server privacy, distributed storage, Lagrange encoding.
\end{IEEEkeywords}

\section{Introduction}
In the modern information age, many emerging technologies, for examples Internet of things, cloud storage and computing, operate on open systems and networks, which brings up several new challenges, particularly, in preserving data privacy and security. Motivated by these challenges there is much interest in the primitives related to preserving privacy and security, among which the problem of Private Information Retrieval (PIR) has drawn remarkable attention in the past few decades. The PIR problem was introduced by Chor \emph{et al.} in \cite{Chor} at first. In the classical PIR setting, a user wishes to retrieve one out of $F$ files from $N$ non-colluding servers, each of which stores all the $F$ files, while preventing any individual server from obtaining information about which file is being requested. To this end, the user prepares $N$ query strings and sends each  to a specific server, such that the designed queries cannot reveal anything about the identity of the desired file to any individual server. Upon receiving the query, each server truthfully responds an answer string with the user according to the information it stores. Finally, the user is able to recover the desired file from the collected answer strings.

The goal of PIR is to design the most efficient solution for the user to retrieve one desired file from a set of distributed servers, without disclosing the identity of the desired file to the servers in the information theoretic security sense. In the original formulation of the PIR problem \cite{single bit3,single bit4,Chor,N_PIR5,single bit2}, each of the files is modeled as one bit size and the communication cost is measured by the total amount of upload cost (the total bit size of query strings) and download cost (the total bit size of answer strings).
A naive strategy for this seemingly impossible task is to download the entire files from the servers no matter which file is requested, which incurs a significant communication cost and therefore is highly inefficient in practice.
It was proved in \cite{Chor} that the naive strategy is the only feasible solution if the files are stored at a single server.
Moreover, it was further shown \cite{Chor} that the communication cost of PIR can be achieved in sublinear scale by replicating the files at multiple non-colluding servers.


Instead of retrieving a single bit, the traditional Shannon theoretic formulation allows the file size to be arbitrary large and accordingly the upload cost can be neglected with respect to the download cost since it does not scale with file size \cite{Ulukus_MDS,code PIR,N. B. Shah,Sun replicated,MDS_Tajeddine2,Tian and Sun,Zhu,Tian_MDS}. The communication effectiveness of a PIR scheme is instead measured by \emph{retrieval rate}, defined as the ratio of the amount of desired information bits to the bit size of download cost. The supremum of PIR rates over all achievable retrieval schemes is referred to as \emph{PIR capacity}.
In the influential work by Sun and Jafar \cite{Sun replicated}, the capacity of classical PIR problem was characterized as $\big(1+\frac{1}{N}+\ldots+\frac{1}{N^{F-1}}\big)^{-1}$ for arbitrary $N$ and $F$.

Since the appearance of \cite{Sun replicated}, a series of works is interested in exploring the fundamental limits of PIR under various variants of the classical model.
This includes PIR with $T$-private queries and replicated storage \cite{Sun_TPIR}, PIR with MDS coded storage \cite{Ulukus_MDS,MDS_Tajeddine2}, RIR with optimal file length \cite{Sun_optimal,Tian and Sun,Zhang_TPIR,Zhu,Tian_MDS}, PIR with $T$-private queries and MDS coded storage \cite{MDS-TPIR,Sun_MDS-TPIR}, PIR with secure storage \cite{X-security,MDS-X-security,Secure:2,Secure_PIR}, PIR with unresponsive and/or Byzantine servers \cite{Ulukus_B-TPIR,MDS-X-security,Tajeddine222}, PIR with arbitrary collusion patterns \cite{arbitray_1,arbitray_2}, multi-round PIR \cite{multi rounds}, multi-message PIR \cite{multi files}, storage constrained PIR \cite{Attia SC-PIR,Tandon Coded caching,Zhu_SCPIR}, PIR with side information \cite{Side_PIR_2,Side_PIR_1,Side_PIR_3}, symmetric PIR \cite{S-PIR1,Wang-MDS-SPIR,wang-colluding-SPIR,wang-byzantine}, as well as extending PIR to private computation \cite{MDS Obead,Raviv PPC,PC1}, private search \cite{private search}, federated submodel learning \cite{Secure:5}, etc.

The work that is most related to ours is \cite{B-PIR} by Jafar \emph{et al.}, which introduces the problem of Multi-user Blind $X$-secure $T$-colluding Symmetric PIR, also referred to as MB-XTSPIR. Specifically, each file, labeled by $M$ indices, is stored in an $X$-secure fashion at $N$ servers, such that any group of up to $X$ colluding servers learn nothing about the files.
There are $M$ users, in which each user $m,m=1,\ldots,M$ privately selects an index $\theta_m$ and wishes to jointly retrieve the file specified by the $M$ user-indices, while keeping its index $\theta_m$ private from any $T_m$ colluding servers.
In addition, each user must not learn any information about the other users' indices and the data files more than the desired one.
The two constraints are called  \emph{blind privacy} and \emph{server privacy} respectively, which are ensured by allowing all the servers to hold some correlated random variables that are independent of the files meanwhile unavailable to the users \cite{S-PIR1,Wang-MDS-SPIR,wang-colluding-SPIR,wang-byzantine,mismatched randomness}, and these variables are referred to as the randomness.
The PIR problem with server privacy is called Symmetric PIR (SPIR).
Naturally, \emph{secrecy rate}, defined as the amount of  randomness relative to the desired file size, becomes another metric to measure the effectiveness of symmetric schemes.
It is conceivable that such a functionality can be directly useful for collecting information cooperatively for multiple intelligence agencies and is useful for multi-way blind medical studies \cite{B-PIR} where the confidential medical data are indexed by multiple attributes, such as the diagnosis, treatment, doctor, health insurance and so on.
The multi-user blind functionality \cite{B-PIR} is also directly useful for a variant of secure multiparty computation \cite{Yao_SMC}. The $M$ users/parties, each holds a private input (user $m$ holds the input $\theta_m$ for all $m=1,\ldots,M$), wish to collectively compute the evaluation of a function $f(x_1,\ldots,x_M)$ at $(\theta_1,\ldots,\theta_M)$, where the file indexed by $(\theta_1,\ldots,\theta_M)$ is the evaluation of the function at $(\theta_1,\ldots,\theta_M)$ and is stored at distributed servers.


As the literature on PIR with various variants continues to grow, it is very desired to find unified perspectives that combine various practical constraints on PIR and establish the essential relationship between these constraints. Particularly, a series of important concern in distributed storage systems is that repetition coding results in extremely large storage cost, and there exists some Byzantine servers that may return arbitrary responses maliciously to prevent the users from retrieving the desired file and unresponsive servers that will not respond any information at all.
Towards this goal, in this paper, we generalize the problem of MB-XTSPIR and focus on U-B-MDS-MB-XTSPIR, i.e., MB-XTSPIR with $(N,K)$ MDS coded storage, $B$ Byzantine servers and $U$ unresponsive servers.
Notably, in order to guarantee blind privacy and server privacy, we consider the case that the randomness stored over all the servers is constructed in the same structure as the outputs of the function
implemented by servers with the queries and stored coded files as inputs, i.e., the coded structure of the function outputs and randomness is matched.\footnote{Surprisingly, it is shown in \cite{mismatched randomness,wang-colluding-SPIR} that the retrieval rate $1-\frac{K}{N}$ of MDS-SPIR (Symmetric PIR with MDS coded storage) \cite{Wang-MDS-SPIR} can be further improved if the coded structure of the function outputs and randomness is mismatched, such as the function outputs are structured in MDS codes while the randomness is replicated. It is also a valuable research direction to explore how to use such mismatched randomness to lift the retrieval rate of the problems that include MDS-SPIR as a special case, for examples the problems in \cite{MDS-X-security,Tajeddine222,wang-colluding-SPIR} and this paper.}

The central technical contribution of this paper is an U-B-MDS-MB-XTSPIR scheme based on Lagrange encoding \cite{LCC}.
For the general setting of U-B-MDS-MB-XTSPIR, it may not be efficient to directly extend the current methods \cite{MDS-TPIR,X-security,MDS-X-security,Tajeddine222,B-PIR} to retrieve the desired file, please refer to Section \ref{comparison} for more explanation. To this end, we create a new form of interference alignment to PIR, 
which takes advantage of the structure inspired by the well-known Lagrange interpolation polynomials to construct storage codes and queries in privacy-preserving manners, such that the answers can be viewed as evaluations of  polynomials and accordingly the decoding is completed by interpolating the polynomials.

As a result, this general scheme achieves a retrieval rate of $1-\frac{K+X+T_1+\ldots+T_M+2B-1}{N-U}$ with secrecy rate $\frac{K+X+T_1+\ldots+T_M-1}{ N-(K+X+T_1+\ldots+T_M+2B+U-1)}$ for any number of files,
which generalizes the current optimal schemes as special cases of U-B-MDS-MB-XTSPIR including SPIR \cite{S-PIR1}, MDS-SPIR \cite{Wang-MDS-SPIR}, TSPIR \cite{wang-byzantine}, MDS-TSPIR \cite{wang-colluding-SPIR}, and MB-XTSPIR \cite{B-PIR}.
Further, when server privacy is not considered, the U-B-MDS-MB-XTSPIR scheme automatically yields the asymptotically optimal schemes for various special PIR cases as the number of files approaches infinity, for examples, U-B-MDS-TPIR \cite{Tajeddine222}, obtained by setting $M=1,X=0$, and U-B-MDS-XTPIR \cite{MDS-X-security}, obtained by setting $M=1$.
Remarkably, the PIR schemes in \cite{Tajeddine222,MDS-X-security} and this paper, all require the decoding to be performed multiple times over the server answers.
In each decoding, references \cite{Tajeddine222,MDS-X-security} first need to use the previously decoded desired symbols to eliminate the interference in server answers and then decode desired symbols, i.e.,  the multiple decoding operations must be performed \emph{in sequence}. However, in our general scheme and its yielded schemes, the multiple decoding operations can be carried out \emph{independently and concurrently}, which improves the efficiency of retrieving desired file.

The rest of this paper is organized as follows. In Section \ref{system model},  the U-B-MDS-MB-XTSPIR problem is formally formulated.
In Section \ref{BPIR scheme}, an U-B-MDS-MB-XTSPIR scheme using Lagrange encoding is constructed.
In Section \ref{comparison}, we compare the proposed general scheme to the related schemes through some simple examples.
Finally, the paper is concluded  in Section \ref{conclusion}.

The following notations are used throughout this paper.
\begin{itemize}
  \item Let boldface and cursive capital letters represent matrices and sets, respectively, e.g., $\mathbf{W}$ and $\mathcal{W}$; 
  \item For any positive integers $m,n$ such that $m\leq n$, $[n]$ and $[m:n]$ denote the set $\{1,2,\ldots,n\}$ and $\{m,m+1,\ldots,n\}$, respectively;
  \item Define $A_{\Gamma}$ as $\{A_{\gamma_1},\ldots,A_{\gamma_{m}}\}$ for any index set $\Gamma=\{\gamma_1,\ldots,\gamma_{m}\}\subseteq[n]$;
  \item For a finite set $\mathcal{X}$, $|\mathcal{X}|$ denotes its cardinality.
\end{itemize}

\section{System Model}\label{system model}
Consider a database $\mathcal{W}$ composed of $F\triangleq F_1F_2\ldots F_M$ files, where each file is labeled by $M$ indices and $F_m$ is the maximum value of the $m$-th index for all $m\in[M]$.
 All the files in the database $\mathcal{W}$ can be indexed as
\begin{IEEEeqnarray*}{rCl}
\mathcal{W}\triangleq\left\{\mathbf{W}^{(f_1,\ldots,f_M)}:f_1\in[F_1],\ldots,f_M\in[F_M]\right\}.
\end{IEEEeqnarray*}

We assume that each file is divided into $K$ blocks, and each block is divided into $\lambda$ stripes.\footnote{We divide the file into $K$ blocks because each file is stored at the distributed storage system according to MDS codes and $K$ is a \emph{fixed parameter} of the MDS codes.
Moreover, in order to improve the flexibility of scheme design, each block is further divided into $\lambda$ stripes, such that the user(s) can efficiently retrieve desired data from server answers.
Notice that as is typical in information theory, the file size is arbitrarily large and the coding scheme may freely choose the parameter $\lambda$, i.e., $\lambda$ is a \emph{free parameter} that needs to be carefully chosen to maximize the effectiveness of schemes.
Typically, such partitioning ideas have been widely applied in distributed storage system to reduce the repair bandwidth when repairing failed nodes from some surviving nodes \cite{A. G. Dimakis222,A. G. Dimakis}.
}
WLOG, we represent the file $\mathbf{W}^{(f_1,\ldots,f_M)}$ by a random matrix of dimension $\lambda\times K$, with each entry choosing independently and uniformly  over the finite field $\mathbb{F}_q$ for some prime power $q$, i.e., for any  $f_1\in[F_1],\ldots,f_M\in[F_M]$,
\begin{IEEEeqnarray}{rCl}\label{file symbols}
\mathbf{W}^{(f_1,\ldots,f_M)}=\left[
  \begin{array}{ccc}
    w^{(f_1,\ldots,f_M)}_{1,1} & \ldots & w^{(f_1,\ldots,f_M)}_{1,K}  \\
    \vdots & \ddots & \vdots \\
    w^{(f_1,\ldots,f_M)}_{\lambda,1} & \ldots &w^{(f_1,\ldots,f_M)}_{\lambda,K}\\
\end{array}
\right]. \IEEEeqnarraynumspace
\end{IEEEeqnarray}
Let $L\triangleq\lambda K$ be the number of symbols contained in the file.
The files are independent of each other, i.e.,  
\begin{IEEEeqnarray}{rCl}
H(\mathcal{W})&=&\sum_{f_1\in[F_1],\ldots,f_M\in[F_M]}H(\mathbf{W}^{(f_1,\ldots,f_M)})=FL, \notag
\end{IEEEeqnarray}
where the entropy function $H(\cdot)$ is calculated with logarithm $q$.

The database is stored at a distributed storage system with $N$ servers according to MDS codes over $\mathbb{F}_q$, while keeping its files secure from any group of up to $X$ colluding servers.
Denote the data stored at server $n$ by $\mathcal{Y}_n$ for any $n\in[N]$.
To guarantee such security and MDS property, the ideas of secret-sharing technology \cite{Shamir} had been widely employed to design secure distributed storage systems in information theory \cite{Secure_PIR,MDS-X-security,Raviv PPC}, such that
\begin{itemize}
  \item \textbf{$X$-Security:} Any $X$ servers remain oblivious perfectly to all the files even if they collude, i.e.,
    \begin{IEEEeqnarray}{rCl}\label{X security}
    I(\mathcal{Y}_{\mathcal{X}};\mathcal{W})=0, \quad\forall\,\mathcal{X}\subseteq[N],|\mathcal{X}|=X.
    \end{IEEEeqnarray}
  \item\textbf{MDS Property:} All the files can be reconstructed by connecting to at least $K+X$ servers to tolerate up to $N-K-X$ server failures, i.e.,
    \begin{IEEEeqnarray}{c}\label{reconstruction}
    H(\mathcal{W}|\mathcal{Y}_{\Gamma})=0,\quad\forall\,\Gamma\subseteq[N],|\Gamma|\geq K+X.\IEEEeqnarraynumspace
    \end{IEEEeqnarray}
    The storage at each server is constrained as $\frac{FL}{K}$, which is reduced by a factor of $\frac{1}{K}$ compared to repetition coding storage, i.e.,
    \begin{IEEEeqnarray}{rCl}
    H(\mathcal{Y}_n)=F\lambda, \quad\forall\,n\in[N].\notag
    \end{IEEEeqnarray}
\end{itemize}
It is obvious that the storage system degrades to the classical $(N,K)$ MDS coded setup \cite{Ulukus_MDS,Zhu,Tian_MDS} when $X=0$.

There are $M$ users, specifying one index each,  who want to jointly and privately retrieve a desired file specified by the indices of the $M$ users from the distributed system.
In particular, each user $m, m\in[M]$ privately selects an index $\theta_m$ from $[F_m]$ and wishes to retrieve the file $\mathbf{W}^{(\theta_1,\ldots,\theta_M)}$ specified by the indices $(\theta_1,\ldots,\theta_M)$ from the $N$ servers, while keeping its index $\theta_m$ private from any group of up to $T_m$ colluding servers, called \textit{user privacy}.
In addition, each user must not learn any information about other users' indices and the database beyond the desired file,
which are also referred to as \emph{blind privacy} and \emph{server privacy}, respectively.

To ensure user privacy, we assume each user owns a private randomness, denoted by $\mathcal{Z}_m$ for user $m\in[M]$.
We also assume that each server $n$ stores a random variable $\widetilde{\mathcal{Z}}_n$ that is independent of the files meanwhile unavailable to the users, which is used for guaranteeing blind privacy and server privacy.
Denote the randomness stored over all the servers by $\widetilde{\mathcal{Z}}=(\widetilde{\mathcal{Z}}_1,\widetilde{\mathcal{Z}}_2,\ldots,\widetilde{\mathcal{Z}}_N)$.
The independence between all the entities is formalized as
\begin{IEEEeqnarray}{l}
H(\mathcal{Y}_{[N]},\theta_{[M]},\mathcal{Z}_{[M]},\widetilde{\mathcal{Z}})=H(\mathcal{Y}_{[N]})\notag\\
\quad\quad\quad\quad\quad+\sum\limits_{m\in[M]}H(\theta_m)+\sum\limits_{m\in[M]}H(\mathcal{Z}_m)+H(\widetilde{\mathcal{Z}}).\IEEEeqnarraynumspace\label{independence:entry}
\end{IEEEeqnarray}


To privately retrieve the desired file, each user uses its randomness to generate and send queries to each server. Upon receiving the queries from all the users, each server accordingly responds with answers according to the information available. Consequently, each user can recover the desired file $\mathbf{W}^{(\theta_1,\ldots,\theta_M)}$ from the answers of servers.
Further, we assume the presence of some servers $\mathcal{B}$ of size at most $B$ that pretend to send arbitrary or worst-case answers to confuse the users retrieving the desired file, known as \emph{Byzantine servers}, and another a set of disjoint servers $\mathcal{U}$ of size at most $U$ that do not respond at all, known as \emph{unresponsive servers}. Notice that all the users have no priori knowledge of the identities of Byzantine servers $\mathcal{B}$ and unresponsive servers $\mathcal{U}$, other than knowing the values of $B$ and $U$.

Throughout this paper, we refer to the above operations as a Multi-user Blind $X$-secure $T$-colluding Symmetric PIR scheme from MDS coded storage system with $B$ Byzantine and $U$ unresponsive servers (or U-B-MDS-MB-XTSPIR in short).
Formally, an U-B-MDS-MB-XTSPIR scheme consists of the following phases:
\begin{enumerate}
\item \emph{Query Phase:} User $m$ generates $N$ queries $Q_{[N]}^{m}$ based on its index $\theta_m$ and private randomness $\mathcal{Z}_m$, i.e.,
\begin{IEEEeqnarray}{c}\notag
H(Q_{[N]}^{m}|\theta_m,\mathcal{Z}_m)=0,\quad\forall\,m\in[M],
\end{IEEEeqnarray}
and then sends ${Q}_{n}^{m}$ to server $n$ for any $n\in[N]$.
\item \emph{Answer Phase:} Upon receiving the queries from the $M$ users, each Byzantine server $n\in\mathcal{B}$ overwrites its answer maliciously and will instead send an arbitrary response $A_{n}$ to confuse the users for any $\mathcal{B}\subseteq[N],|\mathcal{B}|\leq B$. The unresponsive servers $\mathcal{U}$ will not respond any information at all, where $\mathcal{U}\subseteq[N],\mathcal{U}\cap\mathcal{B}=\emptyset$ and $|\mathcal{U}|\leq U$.
And the remaining servers $[N]\backslash(\mathcal{B}\cup\mathcal{U})$, known as \emph{authentic servers}, will respond truthfully the answers, which are the determined functions of the received queries and the stored information, i.e., for any $n\in[N]\backslash(\mathcal{B}\cup\mathcal{U})$,
\begin{IEEEeqnarray}{c}\notag
H(A_{n}|\{{Q}_{n}^{m}\}_{m\in[M]},\mathcal{Y}_n,\widetilde{\mathcal{Z}}_n)=0.\IEEEeqnarraynumspace 
\end{IEEEeqnarray}
\item \emph{Decoding Phase:} Each user $m$ must correctly decode the desired file $\mathbf{W}^{(\theta_1,\ldots,\theta_M)}$ from the answers collected from the responsive servers.
\end{enumerate}
Apparently, the U-B-MDS-MB-XTSPIR scheme must satisfy the following three requirements.
\begin{itemize}
\item\textbf{Correctness:} With the queries and received answers, each user must be able to recover the desired file $\mathbf{W}^{(\theta_1,\ldots,\theta_M)}$, i.e., for any $m\in[M]$,
\begin{IEEEeqnarray}{c}\notag
H(\mathbf{W}^{(\theta_1,\ldots,\theta_M)}|A_{[N]\backslash\mathcal{U}},{Q}_{[N]}^{m})=0.
\end{IEEEeqnarray}
\item\textbf{User Privacy:} The index $\theta_m$ of each user $m$ must be hidden from all the queries sent to any $T_m$ colluding servers, i.e., for any $m\in[M]$ and $\mathcal{T}\subseteq[N],|\mathcal{T}|=T_m$,
\begin{IEEEeqnarray}{c}\label{Infor:priva cons}
I({Q}_{\mathcal{T}}^{m};\theta_m)=0.
\end{IEEEeqnarray}
\item\textbf{Blind Privacy and Server Privacy:} Each user must not gain any additional information in regard to other users' indices and the database more than the desired file, i.e., for any $m\in[M]$,
\begin{IEEEeqnarray}{l}\label{Infor:priva cons2}
I\big(A_{[N]\backslash\mathcal{U}},\theta_m,\mathcal{Z}_m;\mathcal{W},\notag\\
\quad\quad\quad\quad\quad\quad\quad\{\theta_{\overline{m}}\}_{\overline{m}\in[M]\backslash\{m\}}|\mathbf{W}^{(\theta_1,\ldots,\theta_M)}\big)=0.\IEEEeqnarraynumspace
\end{IEEEeqnarray}
\end{itemize}

The performance of an U-B-MDS-MB-XTSPIR scheme can be measured by the following three quantities:\footnote{Notably, with respect to the private randomness $\mathcal{Z}_m$, it can be generated by user $m$ locally and thus  is unnecessary to be considered as a performance metric.}
\begin{enumerate}
  \item[1.] The retrieval rate, which is the number of bits of the desired file that each user can privately retrieve per bit of downloaded data, defined as
  \begin{IEEEeqnarray}{c}\label{rate:1}
  R\triangleq\frac{H(\mathbf{W}^{(\theta_1,\ldots,\theta_M)})}{\sum_{n\in[N]\backslash\mathcal{U}}H(A_{n})}=\frac{L}{D},
  \end{IEEEeqnarray}
    where $D\triangleq\sum_{n\in[N]\backslash\mathcal{U}}H(A_{n})$ is the average download cost from the responsive servers.
  \item[2.] The secrecy rate, which is the amount of the randomness stored at the servers relative to the size of desired file, defined as
  \begin{IEEEeqnarray}{c}\label{rate:2}
  \rho\triangleq\frac{H(\widetilde{\mathcal{Z}})}{H(\mathbf{W}^{(\theta_1,\ldots,\theta_M)})}.
  \end{IEEEeqnarray}
  \item[3.] The finite field size $q$, which ensures the achievability of the coded schemes.
  \end{enumerate}

In principle, the retrieval rate should be maximized across all the U-B-MDS-MB-XTSPIR schemes while the secrecy rate and finite field size are preferred to be as small as possible.



\begin{Remark}\label{general case}
The U-B-MDS-MB-XTSPIR problem in this paper includes as special cases the settings of U-B-MDS-TPIR \cite{Tajeddine222}, obtained by setting $M=1,X=0$ and eliminating server privacy, U-B-MDS-XTPIR \cite{MDS-X-security}, obtained by setting $M=1$ and eliminating server privacy, and MB-XTSPIR \cite{B-PIR}, obtained by setting $K=1,B=U=0$.
\end{Remark}

For clarity, the parameters used in our U-B-MDS-MB-XTSPIR system are listed in Table \ref{tab:parameters}.

\begin{table*}[htbp]
\extrarowheight=4pt
\centering
\caption{Parameters Used in U-B-MDS-MB-XTSPIR}
  \begin{tabular}{|c|c||c|c|}
  \hline
  $\mathcal{W}$ & database & $N$ & number of servers \\ 
  \hline
  $F$ & number of files & $f_m$  & $m$-th index of each file \\ 
  \hline
  $F_m$  & maximum value of the $m$-th index of each file &  $M$ & number of indices specifying each file and number of users \\
  \hline
  $q$ & finite field size & $w_{i,j}^{(f_1,\ldots,f_M)}$ & symbol in row $i$ and column $j$ of file $\mathbf{W}^{(f_1,\ldots,f_M)}$ \\
  \hline
  $\lambda$ & number of rows of each file  & $K$ & number of columns of each file  \\
  \hline
  $\mathcal{Y}_n$ & data stored at server $n$  & $\theta_m$ & index specified by user $m$ \\
  \hline
  $X$ & number of colluding data-curious servers & $T_m$ & number of colluding index-curious servers for user $m$ \\
  \hline
  $\mathcal{Z}_m$ & private randomness owned by user $m$ & $\widetilde{\mathcal{Z}}_n$ & randomness stored by server $n$ \\
  \hline
  $B$ & number of Byzantine servers & $U$ & number of unresponsive servers \\
  \hline
  $Q_{n}^{m}$  & query sent by user $m$ to server $n$ & $A_n$ & answer of server $n$ \\
  \hline
  $R$ & retrieval rate & $\rho$ & secrecy rate \\
  \hline
  \end{tabular}
  \label{tab:parameters}
\end{table*}

\section{U-B-MDS-MB-XTSPIR Scheme Based on Lagrange Encoding}\label{BPIR scheme}
In this section, we present an U-B-MDS-MB-XTSPIR scheme based on Lagrange encoding.

Before that, three useful lemmas are introduced, which will be employed by the scheme to preserve/resist $X$-security, Byzantine and unresponsiveness, and user privacy.

\begin{Lemma}[Generalized Shamir's secret sharing \cite{Shamir}]\label{lemma:security}
Given any positive integers $N,K,X$ such that $N\geq K+X$, let $w_1,\ldots,w_K\in\mathbb{F}_q$ be $K$ secrets and $z_1,\ldots,z_X$ be $X$ random noises distributed independently and uniformly on $\mathbb{F}_q$. Let $\alpha_1,\ldots,\alpha_N$ be $N$ distinct elements from $\mathbb{F}_q$. Let $h_1(\alpha),\ldots,h_K(\alpha),c_1(\alpha),\ldots,c_X(\alpha)$ be arbitrary functions of $\alpha$, then denote
\begin{IEEEeqnarray}{l}
\varphi(\alpha)={w}_{1}h_1(\alpha)+\ldots+{w}_{K}h_K(\alpha)\notag\\
\quad\quad\quad\quad\quad\quad\quad\quad+{z}_{1}c_1(\alpha)+\ldots+{z}_{X}c_X(\alpha). \notag
\end{IEEEeqnarray}
If the matrix
\begin{IEEEeqnarray}{c}\notag
\mathbf{C}=
\left[
  \begin{array}{cccc}
    c_1(\alpha_{n_1}) & c_2(\alpha_{n_1}) & \ldots & c_X(\alpha_{n_1}) \\
    c_1(\alpha_{n_2}) & c_2(\alpha_{n_2}) & \ldots & c_X(\alpha_{n_2}) \\
    \vdots & \vdots & \ddots & \vdots \\
    c_1(\alpha_{n_X}) & c_2(\alpha_{n_X}) & \ldots & c_X(\alpha_{n_X}) \\
  \end{array}
\right]_{X\times X}
\end{IEEEeqnarray}
is non-singular over $\mathbb{F}_q$ for any given $\mathcal{X}=\{n_1,\ldots,n_X\}\subseteq[N]$ with $|\mathcal{X}|=X$, then the $X$ values $\{{\varphi}(\alpha_{n_1}),\ldots,{\varphi}(\alpha_{n_X})\}$  can not learn any information about the $K$ secrets ${w}_{1},\ldots,w_{K}$, i.e., 
\begin{IEEEeqnarray}{c}\notag
I({\varphi}(\alpha_{n_1}),\ldots,{\varphi}(\alpha_{n_X});{w}_{1},\ldots,{w}_{K})=0.
\end{IEEEeqnarray}
\end{Lemma}
\begin{Remark}
Shamir's secret sharing technology \cite{Shamir} tells us that, to divide a secret $w_1$ into $N$ pieces $D_1,\ldots,D_N$ such that
knowledge of any $X$ or fewer pieces can not obtain any information about the secret $w_1$, it is enough to construct a random polynomial $\varphi(\alpha)={w}_{1}+{z}_{1}\alpha+\ldots+{z}_{X}\alpha^X$ and then evaluate $D_n=\varphi(\alpha_n)$ for all $n\in[N]$. Apparently, the technology in Lemma \ref{lemma:security} is a generalization of Shamir's secret sharing \cite{Shamir}, obtained by setting $K=1,h_1(\alpha)=1$ and $c_x(\alpha)=\alpha^x$ for $x\in[X]$.
\end{Remark}

\begin{Lemma}[\cite{Lin}]\label{property:codes}
An $(n,k)$ maximum distance separable code with dimension $k$ and length $n$ is capable against $b$ Byzantine errors and $u$ unresponsive errors if $d_{\min}=n-k+1\geq 2b+u+1$.
\end{Lemma}

\begin{Lemma}[Generalized Cauchy Matrix \cite{Lin}]\label{g-cauchy matrix}
Let $\alpha_1,\ldots,\alpha_k$ and $\beta_1,\ldots,\beta_k$ be the elements from $\mathbb{F}_q$ satisfying $\alpha_i\neq\alpha_j,\beta_i\neq\beta_j$ for any $i\neq j$ and $i,j\in[k]$.
Denote a polynomial of degree $k-1$ by
\begin{IEEEeqnarray}{c}\notag
g_i(\alpha)=\prod\limits_{\ell\in[k]\backslash\{i\}}\frac{\alpha-\beta_\ell}{\beta_i-\beta_\ell},\quad\forall\,i\in[k].
\end{IEEEeqnarray}
Then the following generalized Cauchy matrix $\mathbf{G}$ is invertible over $\mathbb{F}_q$.
\begin{IEEEeqnarray}{c}\notag
\mathbf{G}=
\left[
  \begin{array}{cccc}
    g_{1}(\alpha_1) & g_{2}(\alpha_1)& \ldots & g_{k}(\alpha_1)  \\
    g_{1}(\alpha_2) & g_{2}(\alpha_2)& \ldots & g_{k}(\alpha_2)  \\
    \vdots & \vdots & \ddots & \vdots \\
    g_{1}(\alpha_k) & g_{2}(\alpha_k)& \ldots & g_{k}(\alpha_k)  \\
\end{array}
\right]\cdot\mathrm{diag}(\mathbf{v}),
\end{IEEEeqnarray}
where $\mathbf{v}$ is a vector of dimension $k$ with all entries being non-zero.
\end{Lemma}


In our U-B-MDS-MB-XTSPIR scheme, to retrieve the desired file, user $m,m\in[M]$ generates and sends $S$ queries
\begin{IEEEeqnarray}{c}\label{query:general}
Q_{n}^{m}=(\mathcal{Q}_{n}^{m,1},\mathcal{Q}_{n}^{m,2},\ldots,\mathcal{Q}_{n}^{m,S})
\end{IEEEeqnarray}
to server $n$ for all $n\in[N]$, where the value of $S$ is set to be $K$ and will be explained in the following paragraphs. Upon receiving the queries from all the users, server $n$ accordingly responds with $S$ answers
\begin{IEEEeqnarray}{c}\label{answers:gene}
A_n=(A_n^1,A_n^2,\ldots,A_n^S).
\end{IEEEeqnarray}
In decoding phase, each user first decodes some desired symbols from the collected answers $(A_1^s,A_2^s,\ldots,A_N^s)$ for each $s\in[S]$, and then with the decoded data over all $s=1,2,\ldots,S$, each user is able to recover the desired file $\mathbf{W}^{(\theta_1,\ldots,\theta_M)}$.
For convenience, in line with previous work on U-B-MDS-TPIR/U-B-MDS-XTPIR \cite{Tajeddine222,MDS-X-security}, we refer to each of the $S$ queries and its corresponding answers and decoding as a round, i.e., in our scheme, each round $s,s\in[S]$ means that each user $m$ sends a query $\mathcal{Q}_{n}^{m,s}$ to server $n$, then the server responds an answer $A_n^s$ according to the queries $\{\mathcal{Q}_{n}^{m,s}\}_{m\in[M]}$ received from the $M$ users, and finally each user decodes some desired symbols from the received answers $(A_1^s,A_2^s,\ldots,A_N^s)$.


In each round, in order to efficiently resist Byzantine errors and unresponsive errors such that each user can maximally retrieve the desired data symbols from the answers, it is desirable to enable that the responses of all the servers constitute an MDS codeword because it has maximum code distance.
Intuitively, among the server responses of $N$ dimensions in each round, our scheme exploits $T_1+\ldots+T_M$ dimensions to protect the $M$ user privacies and $K+X-1$ dimensions to eliminate the randomness incurred by the $(N,K+X)$ secure MDS coded data storage. 
In addition, $2B+U$ dimensions are used to correct the $B$ Byzantine errors and $U$ unresponsive errors.
Accordingly, the remaining $N-(K+X+T_1+\ldots+T_M+2B+U-1)$ dimensions are left for each user to retrieve desired data.


Given any U-B-MDS-MB-XTSPIR scheme, let $P$ denote the number of desired data symbols that each user can  privately retrieve in each round of the scheme. In our PIR scheme, the server responses of the remaining $N-(K+X+T+2B+U-1)$ dimensions in each round are completely exploited to retrieve desired data symbols, i.e., our scheme sets
\begin{IEEEeqnarray}{c}
P\triangleq N-(K+X+T_1+\ldots+T_M+2B+U-1)\IEEEeqnarraynumspace \label{improtant para}
\end{IEEEeqnarray}
with $N>K+X+T+2B+U-1$.
Recall that $L=\lambda K$ represents the total number of data symbols that each user needs to retrieve.
The parameter $\lambda$ and number of rounds $S$ should satisfy
\begin{IEEEeqnarray}{c}\notag
PS=\lambda K.
\end{IEEEeqnarray}
In our proposed scheme,  we will set
\begin{IEEEeqnarray}{c}\label{chose:parameters}
\lambda=P,\quad S=K.
\end{IEEEeqnarray}

\subsection{Public  Parameters}\label{Achievable Parameters}
To complete the construction of U-B-MDS-MB-XTSPIR scheme, we first need to generate a group of parameters $\{\beta_{i,j}:i\in[\lambda],j\in[K+X]\}$ and $\{\alpha_1,\ldots,\alpha_N\}$ from $\mathbb{F}_q$. Remarkably, these parameters  are publicized to all the $M$ users and $N$ severs in advance.

For clarity, denote $\{\beta_{i,j}:i\in[\lambda],j\in[K+X]\}$ by a matrix $\boldsymbol{\beta}$ of dimension $\lambda\times(K+X)$,\footnote{Notably, we represent the elements $\{\beta_{i,j}:i\in[\lambda],j\in[K+X]\}$ in matrix form only to understand the following constraints P1-P4 more intuitively.} i.e.,
\begin{IEEEeqnarray}{rCl}\notag
\boldsymbol{\beta}\triangleq\left[
  \begin{array}{ccc;{2pt/2pt}ccc}
    \beta_{1,1} & \ldots & \beta_{1,K} & \beta_{1,K+1} & \ldots & \beta_{1,K+X}  \\
    \vdots & \ddots & \vdots & \vdots & \ddots & \vdots \\
    \beta_{\lambda,1} & \ldots &\beta_{\lambda,K} & \beta_{\lambda,K+1} & \ldots &\beta_{\lambda,K+X} \\
\end{array}
\right].
\end{IEEEeqnarray}
Throughout this paper, let $\{\beta_{i,j},\alpha_n:i\in[\lambda],j\in[K+X],n\in[N]\}\subseteq\mathbb{F}_q$  satisfy
\begin{enumerate}
  \item[P1.] The entries in each row of the matrix $\boldsymbol{\beta}$ are distinct, i.e., given each $i\in[\lambda]$, $\beta_{i,k}\neq\beta_{i,\ell}$ for any $k,\ell\in[K+X]$ such that $k\neq\ell$;
  \item[P2.] For any given $s\in[S]$ where $S=K$, the entries in column $s$ of the matrix $\boldsymbol{\beta}$ are distinct, i.e., $\beta_{k,s}\neq\beta_{\ell,s}$ for any $k,\ell\in[\lambda]$ such that $k\neq\ell$;
  \item[P3.] The elements $\alpha_1,\ldots,\alpha_N$ are distinct, i.e., $\alpha_i\neq\alpha_j$ for any $i,j\in[N]$ such that $i\neq j$;
  \item[P4.] The elements $\alpha_1,\ldots,\alpha_N$ are distinct from the ones in columns $[K]$ of the matrix $\boldsymbol{\beta}$, i.e., $\{\alpha_n:n\in[N]\}\cap\{\beta_{i,j}:i\in[\lambda],j\in[K]\}=\emptyset$.
\end{enumerate}


The following lemma characterizes a lower bound on the size of the finite field $\mathbb{F}_q$ that promises the existence of the parameters $\{\beta_{i,j},\alpha_n:i\in[\lambda],j\in[K+X],n\in[N]\}\subseteq\mathbb{F}_q$ satisfying the above conditions.

\begin{Lemma}[\cite{Zhu_PPC}, Lemma 4]\label{theorem:tield size}
There must exist a group of parameters $\{\beta_{i,j},\alpha_n:i\in[\lambda],j\in[K+X],n\in[N]\}\subseteq\mathbb{F}_q$ satisfying P1-P4 if $q\geq N+\max\{K, \lambda\}$.
\end{Lemma}

Our U-B-MDS-MB-XTSPIR scheme employs the well-known Lagrange interpolation polynomials to create storage codes and queries in privacy-preserving manners. In each round, the answers can be viewed as evaluations of a polynomial, and accordingly the decoding is completed by interpolating the polynomial. This creates a new form of interference alignment to PIR, based on the structure inspired by Lagrange interpolation polynomials.
Specifically, in distributed storage, the key idea is to encode the data files using Lagrange interpolation polynomials of degree $K+X-1$ in a security-preserving manner, and then the evaluations of the Lagrange polynomials at $N$ distinct points are stored at the $N$ distributed servers.
Similarly, in the query phase, the key idea is that user $m,m\in[M]$ uses Lagrange interpolation polynomials of degree $T_m$ to create interference alignment in a privacy-preserving manner, such that the queries sent by user $m$ can be used for eliminating the interference from the files whose $m$-th index is not $\theta_m$. Thus, upon the queries from the $M$ users, the interference from all the undesired files (i.e., all the files except the desired one $\mathbf{W}^{(\theta_1,\ldots,\theta_M)}$) will be eliminated.
In addition, to guarantee that each user can retrieve $P=\lambda$ symbols of the desired file during each round, it is necessary to eliminate the interference between these desired symbols, which is achieved by creating Lagrange polynomials of degree $\lambda-1$, also referred to as intermediate polynomials.
The server answers amount to evaluations of the product polynomial that are consisted of the products of the query polynomials of degree $T_m$ for all the users $m\in[M]$, intermediate polynomials of degree $\lambda-1$, and storage polynomials of degree $K+X-1$.
Accordingly, even though there exists $B$ Byzantine servers and $U$ unresponsive servers, each user can interpolate the product polynomial from the answers collected from the $N$ servers by \eqref{improtant para}, and then recover $\lambda$ desired symbols by evaluating the product polynomial in each round.

In addition, to guarantee blind privacy and server privacy under the constraint of correctness, we use the idea of noise alignment \cite{NA:1,NA:2,Zhu_SDMM} to create noise polynomials that are structured in the same manner as the product polynomial, such that the desired terms in the product polynomial are aligned with zero elements, which preserves correctness, and the remaining interference in the product polynomial that may leak privacy information to users are perfectly masked with random noises.

Next, we describe the data encoding procedures. Then, we present a simple example to illustrate the basic ideas of interference alignment above behind our scheme. Finally, the general U-B-MDS-MB-XTSPIR scheme is formally constructed and its complexity is analysed.

\subsection{Secure Lagrange Storage Codes}\label{LSC}
In this subsection, data encoding procedures are described, where Lagrange interpolation polynomials are used to securely encode each row data of each file separately. For any $f_1\in[F_1],\ldots,f_M\in[F_M]$ and $i\in[\lambda]$, let $z_{i,K+1}^{(f_1,\ldots,f_M)},z_{i,K+2}^{(f_1,\ldots,f_M)},\ldots,z_{i,K+X}^{(f_1,\ldots,f_M)}$ be $X$ random variables distributed independently and uniformly on $\mathbb{F}_q$  
and then choose a polynomial $\varphi_i^{(f_1,\ldots,f_M)}(\alpha)$ of degree at most $K+X-1$ such that
\begin{IEEEeqnarray}{l}
\varphi_i^{(f_1,\ldots,f_M)}(\beta_{i,j})=\notag\\
\quad\quad\quad\quad\quad\left\{
\begin{array}{@{}ll}
w_{i,j}^{(f_1,\ldots,f_M)},&\forall\, j\in[K]\\
z_{i,j}^{(f_1,\ldots,f_M)},&\forall\, j\in[K+1:K+X]
\end{array}\right., \IEEEeqnarraynumspace\label{encdoing}
\end{IEEEeqnarray}
where $w_{i,j}^{(f_1,\ldots,f_M)}$ is the $i$-th element in the $j$-th column of data file $\mathbf{W}^{(f_1,\ldots,f_M)}$ defined in \eqref{file symbols}.

From P1, the Lagrange interpolation rules and the degree restriction guarantee the existence and uniqueness of $\varphi_i^{(f_1,\ldots,f_M)}(\alpha)$, which is written as
\begin{IEEEeqnarray}{c}
\varphi_i^{(f_1,\ldots,f_M)}(\alpha)=\sum\limits_{\ell=1}^{K}w_{i,\ell}^{(f_1,\ldots,f_M)}\cdot\prod_{k\in[K+X]\backslash\{\ell\}}\frac{\alpha-\beta_{i,k}}{\beta_{i,\ell}-\beta_{i,k}}\quad\notag\\
\quad\quad\quad\quad\quad\quad+\sum\limits_{\ell=K+1}^{K+X}z_{i,\ell}^{(f_1,\ldots,f_M)}\cdot\prod_{k\in[K+X]\backslash\{\ell\}}\frac{\alpha-\beta_{i,k}}{\beta_{i,\ell}-\beta_{i,k}}.\notag
\end{IEEEeqnarray}
Then the evaluations of $\varphi_i^{(f_1,\ldots,f_M)}(\alpha)$ $(f_1\in[F_1],\ldots,f_M\in[F_M],i\in[\lambda])$ at point $\alpha=\alpha_n$ are distributedly stored at server $n$, i.e., for any $n\in[N]$,
\begin{IEEEeqnarray}{l}
\mathcal{Y}_n=\Big\{\varphi_i^{(f_1,\ldots,f_M)}(\alpha_n):\notag\\
\quad\quad\quad\quad\quad\quad\quad f_1\in[F_1],\ldots,f_M\in[F_M],i\in[\lambda]\Big\}.\IEEEeqnarraynumspace\label{stored data}
\end{IEEEeqnarray}

In particular, such Lagrange encoding is equivalent to the $(N,K+X)$ Reed-Solomon (RS) code \cite{Raviv PPC} with a class of specific basis polynomials $\sigma_{i,1}(\alpha),\sigma_{i,2}(\alpha),\ldots,\sigma_{i,K+X}(\alpha)$ for any $i\in[\lambda]$, where
\begin{IEEEeqnarray}{c}\notag
\sigma_{i,\ell}(\alpha)=\prod_{k\in[K+X]\backslash\{\ell\}}\frac{\alpha-\beta_{i,k}}{\beta_{i,\ell}-\beta_{i,k}},\quad\forall\,\ell\in[K+X].
\end{IEEEeqnarray}
By P3, $\big(\varphi_i^{(f_1,\ldots,f_M)}(\alpha_1),\ldots,\varphi_i^{(f_1,\ldots,f_M)}(\alpha_N)\big)$ is an $(N,K+X)$ RS codeword over $\mathbb{F}_q$ for any $f_1\in[F_1],\ldots,f_M\in[F_M],i\in[\lambda]$ and thus such Lagrange storage encoding has the property of $(N,K+X)$ MDS codes.


\subsection{Example for Illustration}\label{example:scheme}
In this subsection, we illustrate our scheme through a simple example for the parameters $N=13,M=2,K=2,X=2,T_1=2,T_2=2,B=1,U=1$, which induce $\lambda=P=3$ and $S=2$.

Here, each file $\mathbf{W}^{(f_1,f_2)}$ for any $f_1\in[F_1],f_2\in[F_2]$ is the form of
\begin{IEEEeqnarray}{rCl}\label{exam:file symbols}
\mathbf{W}^{(f_1,f_2)} =\left[
  \begin{array}{cc}
    w^{(f_1,f_2)}_{1,1} & w^{(f_1,f_2)}_{1,2}  \\
    w^{(f_1,f_2)}_{2,1} & w^{(f_1,f_2)}_{2,2}\\
    w^{(f_1,f_2)}_{3,1} & w^{(f_1,f_2)}_{3,2}\\
\end{array}
\right].
\end{IEEEeqnarray}

\subsubsection*{Lagrange Data Encoding}
Let $\{\beta_{i,j},\alpha_n:i\in[3],j\in[4],n\in[13]\}\subseteq\mathbb{F}_q$ be a set of parameters satisfying P1-P4. For every $f_1\in[F_1],f_2\in[F_2]$ and $i\in[3]$, choose $X=2$ random variables $z_{i,3}^{(f_1,f_2)},z_{i,4}^{(f_1,f_2)}$ independently and uniformly from $\mathbb{F}_q$, and then design the Lagrange interpolation polynomial $\varphi_i^{(f_1,f_2)}(\alpha)$ of degree $K+X-1=3$ such that
\begin{IEEEeqnarray}{rClrCl}
\varphi_i^{(f_1,f_2)}(\beta_{i,1})&=&w_{i,1}^{(f_1,f_2)},\quad
\varphi_i^{(f_1,f_2)}(\beta_{i,2})&=&w_{i,2}^{(f_1,f_2)},\IEEEeqnarraynumspace\label{example:storage}\\
\varphi_i^{(f_1,f_2)}(\beta_{i,3})&=&z_{i,3}^{(f_1,f_2)},\quad
\varphi_i^{(f_1,f_2)}(\beta_{i,4})&=&z_{i,4}^{(f_1,f_2)}. \IEEEeqnarraynumspace\notag
\end{IEEEeqnarray}
The data stored at server $n\in[13]$ is given by
\begin{IEEEeqnarray}{c}\notag
\mathcal{Y}_n=\left\{\varphi_i^{(f_1,f_2)}(\alpha_n):f_1\in[F_1],f_2\in[F_2],i\in[3]\right\}.
\end{IEEEeqnarray}

\subsubsection*{U-B-MDS-MB-XTSPIR Scheme}
The scheme will happen over $S=2$ rounds, and in round $s\in[2]$ each user can decode the $P=\lambda=3$ symbols in the $s$-th column of the desired file $\mathbf{W}^{(\theta_1,\theta_2)}$.

During round $s\in[2]$, each user $m\in[2]$ independently and uniformly generates $F_m\lambda T_m=6F_m$ random variables
\begin{IEEEeqnarray}{l}
\mathcal{Z}_m^{s}=\Big\{z_{j,t}^{(f_m),m,s}: f_m\in[F_m],j\in[3],t\in[2]\Big\} \notag
\end{IEEEeqnarray}
from $\mathbb{F}_q$, and then sends the following query $\mathcal{Q}_{n}^{m,s}$ to server $n$ for any $n\in[13]$:
\begin{IEEEeqnarray}{l}
\mathcal{Q}_{n}^{m,s}=\Big\{Q_{j}^{(f_m),m,s}(\alpha_n): f_m\in[F_m],j\in[3]\Big\},\IEEEeqnarraynumspace \label{query:example}
\end{IEEEeqnarray}
where $Q^{(f_m),m,s}_{j}(\alpha)$ is a Lagrange polynomial of degree $T_m=2$ such that
\begin{IEEEeqnarray}{rCl}
Q^{(f_m),m,s}_{j}(\beta_{j,s})&=&\left\{
\begin{array}{@{}ll}
1, &\mathrm{if}\,\,f_m=\theta_m\\
0, &\mathrm{if}\,\, f_m\neq\theta_m
\end{array}
\right.,\label{example:answer111}\\
Q^{(f_m),m,s}_{j}(\alpha_1)&=&z_{j,1}^{(f_m),m,s},\notag\\
Q^{(f_m),m,s}_{j}(\alpha_2)&=&z_{j,2}^{(f_m),m,s}, \notag
\end{IEEEeqnarray}
which ensure user privacy and are used to eliminate the interference from the files whose $m$-th index is not $\theta_m$.

Before responding the queries, each server constructs three intermediate polynomials $\phi_1^s(\alpha),\phi_2^s(\alpha),\phi_3^s(\alpha)$ of degree $\lambda-1=2$ such that
\begin{IEEEeqnarray}{rClrClrCl}
\phi_1^s(\beta_{1,s})&=&1, \quad \phi_1^s(\beta_{2,s})&=&0, \quad \phi_1^s(\beta_{3,s})&=&0, \IEEEeqnarraynumspace\label{example111} \\
\phi_2^s(\beta_{1,s})&=&0, \quad \phi_2^s(\beta_{2,s})&=&1, \quad \phi_2^s(\beta_{3,s})&=&0, \IEEEeqnarraynumspace \\
\phi_3^s(\beta_{1,s})&=&0, \quad \phi_3^s(\beta_{2,s})&=&0, \quad \phi_3^s(\beta_{3,s})&=&1, \IEEEeqnarraynumspace\label{example222}
\end{IEEEeqnarray}
which are used to eliminate the interference between the $\lambda=3$ symbols in the $s$-th column of each file.

Moreover, to protect blind privacy and server privacy, let $z_{1}^{s},\ldots,z_{7}^{s}$ be the random variables that are unavailable to the users.
Then, the randomness $\widetilde{\mathcal{Z}}_n^s$ stored by server $n$ is given by
\begin{IEEEeqnarray}{c}\label{stored randomness:exam}
\widetilde{\mathcal{Z}}_n^s=\psi^s(\alpha_n),
\end{IEEEeqnarray}
where $\psi^s(\alpha)$ is a noise polynomial of degree $\lambda+K+X+T_1+T_2-2=9$ such that
\begin{IEEEeqnarray}{l}
\psi^s(\beta_{1,s})=0,\,\, \psi^s(\beta_{2,s})=0, \,\, \psi^s(\beta_{3,s})=0,  \IEEEeqnarraynumspace\label{decoding:example}\\
\psi^s(\alpha_{1})=z_{1}^{s}, \,\, \psi^s(\alpha_{2})=z_{2}^{s}, \,\, \psi^s(\alpha_{3})=z_{3}^{s}, \IEEEeqnarraynumspace\\
\psi^s(\alpha_{4})=z_{4}^{s}, \,\, \psi^s(\alpha_{5})=z_{5}^{s}, \,\, \psi^s(\alpha_{6})=z_{6}^{s},\,\,  \psi^s(\alpha_{7})=z_{7}^{s}. \IEEEeqnarraynumspace\label{anwerasd;exam2}
\end{IEEEeqnarray}


By employing the received queries $\{\mathcal{Q}_{n}^{m,s}\}_{m\in[2]}$ from the $M=2$ users and the intermediate polynomials $\phi_1^s(\alpha),\phi_2^s(\alpha),\phi_3^s(\alpha)$ as coefficients, server $n$ computes a linear combination of the stored encoded symbols $\mathcal{Y}_n$, and then responds the sum of the linear combination and the randomness $\widetilde{\mathcal{Z}}_n^s=\psi^s(\alpha_n)$ stored by server $n$ for the users:
\begin{IEEEeqnarray}{l}
A^s_n=\sum\limits_{f_1\in[F_1]}\sum\limits_{f_2\in[F_2]}\sum\limits_{j\in[3]}\phi_j^s(\alpha_n)\cdot Q^{(f_1),1,s}_{j}(\alpha_n)\notag\\
\quad\quad\quad\quad\quad\quad\times Q^{(f_2),2,s}_{j}(\alpha_n)\cdot \varphi_j^{(f_1,f_2)}(\alpha_n)+\psi^s(\alpha_n).\notag 
\end{IEEEeqnarray}

Denote by $A^s(\alpha)$ the answer polynomial of degree $\lambda+K+X+T_1+T_2-2=9$, where
\begin{IEEEeqnarray}{l}
A^s(\alpha)=\sum\limits_{f_1\in[F_1]}\sum\limits_{f_2\in[F_2]}\sum\limits_{j\in[3]}\phi_j^s(\alpha)\cdot Q^{(f_1),1,s}_{j}(\alpha)\notag\\
\quad\quad\quad\quad\quad\quad\quad \times Q^{(f_2),2,s}_{j}(\alpha)\cdot \varphi_j^{(f_1,f_2)}(\alpha)+\psi^s(\alpha).\notag 
\end{IEEEeqnarray}
That is, the response $A_n^s$ of server $n$ is equivalent to evaluating $A^s(\alpha)$ at $\alpha=\alpha_n$ for any authentic server $n$.
Thus, the answers $(A_1^s,\ldots,A_{13}^s)$ from all the servers form a $(13,10)$ RS codeword, which is robust against any $B=1$ Byzantine error and $U=1$ unresponsive error. Accordingly, each user can recover the polynomial $A^s(\alpha)$ from the answers of servers by using RS decoding algorithms even if there exists $B=1$ Byzantine answer and $U=1$ unresponsive answer. 

Finally, for any $i\in[3]$, each user evaluates the polynomial $A^s(\alpha)$ at $\alpha=\beta_{i,s}$ and obtains
\begin{IEEEeqnarray}{rCl}
A^s(\beta_{i,s})&=&\sum\limits_{f_1\in[F_1]}\sum\limits_{f_2\in[F_2]}\sum\limits_{j\in[3]}\phi_j^s(\beta_{i,s})\cdot Q^{(f_1),1,s}_{j}(\beta_{i,s})\notag\\
&&\times Q^{(f_2),2,s}_{j}(\beta_{i,s})\cdot\varphi_j^{(f_1,f_2)}(\beta_{i,s})+\psi^s(\beta_{i,s}) \IEEEeqnarraynumspace\label{anwerasd;exam}\\
&\overset{(a)}{=}&\sum\limits_{f_1\in[F_1]}\sum\limits_{f_2\in[F_2]}Q^{(f_1),1,s}_{i}(\beta_{i,s})\notag\\
&&\,\,\quad\quad\quad\quad\times Q^{(f_2),2,s}_{i}(\beta_{i,s})\cdot\varphi_i^{(f_1,f_2)}(\beta_{i,s}) \label{answer:poly1234}\\
&\overset{(b)}{=}&\varphi_i^{(\theta_1,\theta_2)}(\beta_{i,s}) \label{answer:poly22}\\
&\overset{(c)}{=}& w_{i,s}^{(\theta_1,\theta_2)},\label{evaluating:123122}
\end{IEEEeqnarray}
where $(a)$ is due to \eqref{example111}-\eqref{example222}  and \eqref{decoding:example}; $(b)$ follows by \eqref{example:answer111}; $(c)$ follows from \eqref{example:storage}.

It is straightforward from the decoding process \eqref{anwerasd;exam}-\eqref{evaluating:123122} to obtain some intuitions why the intermediate polynomials $\phi_1^s(\alpha),\phi_2^s(\alpha),\phi_3^s(\alpha)$, the query polynomial $Q^{(f_m),m,s}_{j}(\alpha)$ and the noise polynomial $\psi^s(\alpha)$ are constructed in this way:
\begin{itemize}
\item The intermediate polynomials $\phi_1^s(\alpha),\phi_2^s(\alpha),\phi_3^s(\alpha)$ in \eqref{example111}-\eqref{example222}  are constructed for eliminating the interference between the symbols $w_{1,s}^{(f_1,f_2)},w_{2,s}^{(f_1,f_2)},w_{3,s}^{(f_1,f_2)}$ in the $s$-th column of the file labeled by the indices $(f_1,f_2)$ for any $f_1\in[F_1]$ and $f_2\in[F_2]$, as shown in \eqref{anwerasd;exam}-\eqref{answer:poly1234}.
\item For the query polynomials $Q^{(f_m),m,s}_{j}(\alpha)$, the constraint in \eqref{example:answer111} ensures that the queries \eqref{query:example} can eliminate the interference from the files whose $m$-th index is not $\theta_m$. Accordingly, upon the queries from the $M=2$ users, the interference from all the undesired files are completely eliminated, see the decoding process in \eqref{answer:poly1234}-\eqref{answer:poly22}.
Furthermore, $Q^{(f_m),m,s}_{j}(\alpha_1)=z_{j,1}^{(f_m),m,s}$ and $Q^{(f_m),m,s}_{j}(\alpha_2)=z_{j,2}^{(f_m),m,s}$ are used for ensuring that
the query elements sent to any $T_m=2$ servers reveal nothing about the index $\theta_m$ even if they collude.
\item The noise polynomial $\psi^s(\alpha)$ satisfying \eqref {decoding:example}-\eqref{anwerasd;exam2} is constructed for reserving the desired symbols $w_{1,s}^{(\theta_1,\theta_2)},w_{2,s}^{(\theta_1,\theta_2)},w_{3,s}^{(\theta_1,\theta_2)}$ in answer polynomial $A^s(\alpha)$ and using random noises $z_{1}^{s},\ldots,z_{7}^{s}$ to mask all the residual interference in $A^s(\alpha)$, such that decodability is kept, and  blind privacy and server privacy are protected.
\end{itemize}

Finally, each user can decode the desired symbols $w_{1,s}^{(\theta_1,\theta_2)},$ $w_{2,s}^{(\theta_1,\theta_2)},w_{3,s}^{(\theta_1,\theta_2)}$ by evaluating $A^s(\alpha)$ at $\alpha=\beta_{1,s},\beta_{2,s},\beta_{3,s}$,
and accordingly recover the desired file $\mathbf{W}^{(\theta_1,\theta_2)}$ \eqref{exam:file symbols} after rounds $s=1,2$.
The scheme achieves the retrieval rate $R=\frac{1}{4}$ with secrecy rate $\rho=\frac{7}{3}$.


\subsection{General Construction for U-B-MDS-MB-XTSPIR Scheme}\label{s-PPC scheme}

To privately retrieve the desired file $\mathbf{W}^{(\theta_1,\ldots,\theta_M)}$, the queries, answers and decoding of $S=K$ rounds will be operated as follows.

In general, during round $s\in[S]$, our scheme enables each user to retrieve the $P=\lambda$ symbols in column $s$ of the desired file $\mathbf{W}^{(\theta_1,\ldots,\theta_M)}$ from the $N-U$ answers of responsive servers.

To ensure user privacy, each user $m\in[M]$ generates independently and uniformly $F_m\lambda T_m$ random variables
\begin{IEEEeqnarray}{l}
\mathcal{Z}_m^s=\Big\{z_{j,t}^{(f_m),m,s}: f_m\in[F_m],j\in[\lambda],t\in[T_m]\Big\} \IEEEeqnarraynumspace\label{user randomness:round}
\end{IEEEeqnarray}
from $\mathbb{F}_q$.
Then, for any given $f_m\in[F_m]$ and $j\in[\lambda]$, user $m$ constructs a query polynomial $Q_j^{(f_m),m,s}(\alpha)$ of degree $T_m$ such that
\begin{IEEEeqnarray}{rClll}
Q_j^{(f_m),m,s}(\beta_{j,s})&=&\left\{
\begin{array}{@{}ll}
1, &\mathrm{if}\,\,f_m=\theta_m\\
0, &\mathrm{if}\,\, f_m\neq\theta_m
\end{array}
\right.,&\label{symmetric:1111}\\
Q_j^{(f_m),m,s}(\alpha_{t})&=&z_{j,t}^{(f_m),m,s},\quad\forall\, t\in[T_m]. \notag
\end{IEEEeqnarray}

It will be shown in decoding process \eqref{query:construct}-\eqref{evaluating:111} that the polynomials $Q_j^{(f_m),m,s}(\alpha)$  are constructed to  eliminate the interference from undesired files whose $m$-th index is not $\theta_m$, while keeping the index $\theta_m$ private for any $T_m$ colluding servers.
By P3-P4, the $T_m+1$ elements $\{\beta_{j,s},\alpha_t:t\in[T_m]\}$ are distinct for any $j\in[\lambda]$ and $s\in[S]$. Thus, $Q_j^{(f_m),m,s}(\alpha)$ can be accurately expressed as
\begin{IEEEeqnarray}{l}
Q_j^{(f_m),m,s}(\alpha)=\notag\\
\quad\quad\sum\limits_{\ell\in[T_m]}z_{j,\ell}^{(f_m),m,s}\cdot\frac{\alpha-\beta_{j,s}}{\alpha_{\ell}-\beta_{j,s}}\cdot\prod\limits_{v\in[T_m]\backslash\{\ell\}}\frac{\alpha-\alpha_v}{\alpha_{\ell}-\alpha_v}\notag\\
\quad\quad\quad\quad\quad\quad\quad\quad+\left\{
\begin{array}{@{}ll}
\prod\limits_{v\in[T_m]}\frac{\alpha-\alpha_v}{\beta_{j,s}-\alpha_v}, &\mathrm{if}\,\,f_m=\theta_m\\
0, &\mathrm{if}\,\, f_m\neq\theta_m
\end{array}
\right.. \IEEEeqnarraynumspace\label{symmetric:answer11}
\end{IEEEeqnarray}


Next, the user $m$ evaluates all the $F_m\lambda$ query polynomials at $\alpha=\alpha_n$ and then sends them to server $n\in[N]$:
\begin{IEEEeqnarray}{l}\label{query:1}
\mathcal{Q}_{n}^{m,s}=\Big\{Q_{j}^{(f_m),m,s}(\alpha_n): f_m\in[F_m],j\in[\lambda]\Big\}.\IEEEeqnarraynumspace
\end{IEEEeqnarray}

Before responding the queries, each server  constructs $\lambda$ intermediate polynomials of degree $\lambda-1$ as
\begin{IEEEeqnarray}{c}\label{answer:11}
\phi_j^{s}(\alpha)=\prod\limits_{k\in[\lambda]\backslash\{j\}}\frac{\alpha-\beta_{k,s}}{\beta_{j,s}-\beta_{k,s}},\quad\forall\,j\in[\lambda],
\end{IEEEeqnarray}
which satisfies
\begin{IEEEeqnarray}{rCll}
\phi_j^{s}(\beta_{i,s})&=&\left\{
\begin{array}{@{}ll}
1, &\mathrm{if}\,\, j=i\\
0, & \mathrm{otherwise}
\end{array}
\right.,& \quad\forall\, i\in[\lambda].\label{answer:22}
\end{IEEEeqnarray}
It can be observed from the decoding process \eqref{query:construct123}-\eqref{query:construct} that the $\lambda$ polynomials are used to eliminate the interference between the $\lambda$ symbols in column $s$ of each file.

Let $\{z_{i}^{s}:i\in[K+X+T_1+\ldots+T_M-1]\}$ be another $K+X+T_1+\ldots+T_M-1$ random variables distributed independently and uniformly over $\mathbb{F}_q$.
To ensure blind privacy and server privacy, construct a noise polynomial $\psi^s(\alpha)$ of degree $\lambda+K+X+T_1+\ldots+T_M-2$ such that
\begin{IEEEeqnarray}{rCll}
\psi^s(\beta_{i,s})&=&0,&\quad\forall\, i\in[\lambda], \label{symmetric:1}\\
\psi^s(\alpha_{i})&=&z_{i}^{s},&\quad\forall\, i\in[K+X+T_1+\ldots+T_M-1]. \IEEEeqnarraynumspace \label{symmetric:122}
\end{IEEEeqnarray}
Recall from P2-P4 that $\beta_{i,s},i\in[\lambda]$ and $\alpha_i,i\in[K+X+T_1+\ldots+T_M-1]$ are pairwise distinct elements from $\mathbb{F}_q$. Thus, the polynomial $\psi^s(\alpha)$ is the form of
\begin{IEEEeqnarray}{l}
\psi^s(\alpha)=\sum\limits_{\ell\in[K+X+T_1+\ldots+T_M-1]}z_{\ell}^{s}\cdot \bigg(\prod\limits_{k\in[\lambda]}\frac{\alpha-\beta_{k,s}}{\alpha_{\ell}-\beta_{k,s}}\bigg)\notag\\
\quad\quad\quad\quad\quad\quad\times\bigg(\prod\limits_{v\in[K+X+T_1+\ldots+T_M-1]\backslash\{\ell\}}\frac{\alpha-\alpha_{v}}{\alpha_{\ell}-\alpha_{v}}\bigg).\IEEEeqnarraynumspace\label{symmetric:answer}
\end{IEEEeqnarray}
Then, the random variable $\widetilde{\mathcal{Z}}_n^s$ stored by server $n$ is given by evaluating the noise polynomial $\psi^s(\alpha)$ at $\alpha=\alpha_n$:
\begin{IEEEeqnarray}{c}\label{stored randomness}
\widetilde{\mathcal{Z}}_n^s=\psi^s(\alpha_n).
\end{IEEEeqnarray}
That is, the randomness $\widetilde{\mathcal{Z}}^s=(\widetilde{\mathcal{Z}}_1^s,\ldots,\widetilde{\mathcal{Z}}_N^s)$ in round $s$ is stored at the $N$ distributed servers according to $(N,\lambda+K+X+T_1+\ldots+T_M-1)$ RS codes.

Upon receiving the queries \eqref{query:1} from the $M$ users, server $n$ computes a response $A_n^s$ for the users, based on the stored data in \eqref{stored data}, the intermediate polynomials in \eqref{answer:11} and the stored random variable in \eqref{stored randomness}:
\begin{IEEEeqnarray}{l}
A_n^s=\sum\limits_{f_1\in[F_1],\ldots,f_M\in[F_M]}\sum\limits_{j\in[\lambda]}\phi_j^{s}(\alpha_n)\notag\\
\times\bigg(\prod\limits_{m\in[M]}Q_{j}^{(f_m),m,s}(\alpha_n)\bigg)\cdot\varphi_j^{(f_1,\ldots,f_M)}(\alpha_n)+\psi^s(\alpha_n). \IEEEeqnarraynumspace\label{answers}
\end{IEEEeqnarray}
Notice that there are at most $B$ Byzantine servers, each of which instead generates an arbitrary element from $\mathbb{F}_q$ to confuse the users.
Meanwhile, there are at most $U$ unresponsive servers that will not respond any information at all.

Denote the answer polynomial by
\begin{IEEEeqnarray}{l}
A^s(\alpha)=\sum\limits_{f_1\in[F_1],\ldots,f_M\in[F_M]}\sum\limits_{j\in[\lambda]}\phi_j^{s}(\alpha)\notag\\
\quad\times\bigg(\prod\limits_{m\in[M]}Q_{j}^{(f_m),m,s}(\alpha)\bigg)\cdot\varphi_j^{(f_1,\ldots,f_M)}(\alpha)+\psi^s(\alpha). \IEEEeqnarraynumspace\label{answer:polynomial}
\end{IEEEeqnarray}

Clearly, the answer $A_n^s$ is equivalent to evaluating $A^s(\alpha)$ at $\alpha=\alpha_n$ for any authentic server $n\in[N]\backslash(\mathcal{B}\cup\mathcal{U})$.
It is straight to prove that the degree of $A^s(\alpha)$ is $\lambda+K+X+T_1+\ldots+T_M-2$ in the variable $\alpha$. From P3 again, $\alpha_1,\ldots,\alpha_N$ are distinct elements from $\mathbb{F}_q$.
Thus, $\big(A^s(\alpha_1),\ldots,A^s(\alpha_N)\big)$ forms an $(N,\lambda+K+X+T_1+\ldots+T_M-1)$ RS codeword, which provides robustness against $B$ random errors and $U$ erasure errors by \eqref{chose:parameters} and Lemma \ref{property:codes}.
Accordingly, each user can decode the polynomial $A^s(\alpha)$ from the answers $(A_1^s,\ldots,A_N^s)$ by using RS decoding algorithms \cite{Lin,Gao}  even if there exists $B$ Byzantine servers and $U$ unresponsive servers.


For any $i\in[\lambda]$, evaluating $A^s(\alpha)$ at $\alpha=\beta_{i,s}$ has
\begin{IEEEeqnarray}{rCl}
A^s(\beta_{i,s})
&=&\sum\limits_{f_1\in[F_1],\ldots,f_M\in[F_M]}\sum\limits_{j\in[\lambda]}\phi_j^{s}(\beta_{i,s})\notag\\
&&\times\bigg(\prod\limits_{m\in[M]}Q_{j}^{(f_m),m,s}(\beta_{i,s})\bigg)\cdot\varphi_j^{(f_1,\ldots,f_M)}(\beta_{i,s})\notag\\
&&\quad\quad\quad\quad\quad\quad\quad\quad\quad\quad\quad\quad+\psi^s(\beta_{i,s}) \label{query:construct123}\\
&\overset{(a)}{=}&\sum\limits_{f_1\in[F_1],\ldots,f_M\in[F_M]}\bigg(\prod\limits_{m\in[M]}Q_{i}^{(f_m),m,s}(\beta_{i,s})\bigg)\notag\\
&&\quad\quad\quad\quad\quad\quad\quad\quad\quad\times\varphi_i^{(f_1,\ldots,f_M)}(\beta_{i,s})\label{query:construct}\\
&\overset{(b)}{=}&\varphi_i^{(\theta_1,\ldots,\theta_M)}(\beta_{i,s}) \label{evaluating:111}\\
&\overset{(c)}{=}& w_{i,s}^{(\theta_1,\ldots,\theta_M)},\label{evaluating:1}
\end{IEEEeqnarray}
where $(a)$ follows from \eqref{answer:22} and \eqref{symmetric:1};
$(b)$ is due to \eqref{symmetric:1111}; $(c)$ is due to \eqref{encdoing}.

By evaluating $A^s(\alpha)$ at $\alpha=\beta_{1,s},\beta_{2,s},\ldots,\beta_{\lambda,s}$, each user $m$ can obtain all the symbols in column $s$ of the desired file $\mathbf{W}^{(\theta,\ldots,\theta_M)}$. As a result, each user recovers the desired file $\mathbf{W}^{(\theta_1,\ldots,\theta_M)}$ \eqref{file symbols} after traversing $s\in[S]$, where $S=K$ by \eqref{chose:parameters}.

The performance of the proposed scheme is characterized in the following theorem.
\begin{Theorem}
If $N>K+X+T_1+\ldots+T_M+2B+U-1$, the proposed U-B-MDS-MB-XTSPIR scheme using Lagrange encoding achieves
\begin{IEEEeqnarray*}{l}
\text{Retrieval Rate:}\\
\quad\quad\quad  R=1-\frac{K+X+T_1+\ldots+T_M+2B-1}{N-U}, \\
\text{Secrecy Rate:}\\
\quad\quad \rho=\frac{K+X+T_1+\ldots+T_M-1}{ N-(K+X+T_1+\ldots+T_M+2B+U-1)}, \\ 
\text{Finite Field Size:}\quad q\geq N+\max\{K, \\
\quad\quad\quad N-(K+X+T_1+\ldots+T_M+2B+U-1)\}.
\end{IEEEeqnarray*}
\end{Theorem}
\begin{IEEEproof}
By Lemma \ref{theorem:tield size},  the finite field $\mathbb{F}_q$ is enough with size $q\geq N+\max\{K, N-(K+X+T_1+\ldots+T_M+2B+U-1)\}$.
In each round, the users download $N-U$ symbols from the responsive servers. Thus, by \eqref{chose:parameters}, the retrieval rate \eqref{rate:1} is
\begin{IEEEeqnarray*}{rCl}
R&=&\frac{\lambda K}{\sum_{s=1}^{S}(N-U)}\\
&=&1-\frac{K+X+T_1+\ldots+T_M+2B-1}{N-U}.
\end{IEEEeqnarray*}
From \eqref{rate:2}, the security rate is given by
\begin{IEEEeqnarray*}{rCl}
\rho&=&\frac{\sum_{s=1}^{S}H(\widetilde{\mathcal{Z}}_1^s,\widetilde{\mathcal{Z}}_2^s,\ldots,\widetilde{\mathcal{Z}}_N^s)}{H(\mathbf{W}^{(\theta_1,\ldots,\theta_M)})} \\
&\overset{(a)}{=}&\frac{\sum_{s=1}^{S}(K+X+T_1+\ldots+T_M-1)}{\lambda K}\notag\\
&=&\frac{K+X+T_1+\ldots+T_M-1}{ N-(K+X+T_1+\ldots+T_M+2B+U-1)},\label{secure:rate}
\end{IEEEeqnarray*}
where $(a)$ follows by \eqref{symmetric:answer} and \eqref{stored randomness}. 

Moreover, $X$-security, user privacy, blind privacy and server-privacy will be proved in Appendix \ref{proof:achevable}.
\end{IEEEproof}
\begin{Remark}
Recall from Remark \ref{general case} that, U-B-MDS-MB-XTSPIR includes as special cases the settings of MB-XTSPIR \cite{B-PIR}, obtained by setting $K=1,B=U=0$, U-B-MDS-XTPIR \cite{MDS-X-security}, obtained by setting $M=1$ and eliminating server privacy, and U-B-MDS-TPIR \cite{Tajeddine222}, obtained by setting $M=1,X=0$ and eliminating server privacy. The performance of these schemes are compared in Table \ref{performance PIR}.
Obviously, the performance of our U-B-MDS-MB-XTSPIR scheme in the corresponding special cases are consistent with that in \cite{B-PIR,MDS-X-security,Tajeddine222}, except slightly increasing the field size compared to the U-B-MDS-XTPIR scheme \cite{MDS-X-security} and the U-B-MDS-TPIR scheme \cite{Tajeddine222}.
\begin{table*}[htbp]
\renewcommand{\arraystretch}{1.25}
\centering
\caption{Performance comparison for the schemes in \cite{B-PIR,MDS-X-security,Tajeddine222} and this paper} \label{performance PIR}
\begin{threeparttable}
  \begin{tabular}{|c|c|c|c|}
  \hline
   & Retrieval Rate & Secrecy Rate & Finite Field Size\\ \hline
  U-B-MDS-MB-XTSPIR  & $1-\frac{K+X+T+2B-1}{N-U}$ & $\frac{K+X+T-1}{ N-(K+X+T+2B+U-1)}$ & $q\geq N+\max\{K, N-(K+X+T+2B+U-1)\}$\\ \hline
 MB-XTSPIR \cite{B-PIR} & $1-\frac{X+T}{N}$ & $\frac{X+T}{ N-(X+T)}$ & $q\geq 2N-(X+T)$ \\ \hline
  U-B-MDS-XTPIR \cite{MDS-X-security} & $1-\frac{K+X+T_1+2B-1}{N-U}$ & \cmarkk & $q\geq 2N-(K+X+T_1+2B+U-1)$ \\ \hline
  U-B-MDS-TPIR \cite{Tajeddine222} & $1-\frac{K+T_1+2B-1}{N-U}$ & \cmarkk & $q\geq N$ \\ \hline
  \end{tabular}
  \begin{tablenotes}
       \footnotesize
       \item[]Here, $T=T_1+\ldots+T_M$.
     \end{tablenotes}
\end{threeparttable}
\end{table*}
\end{Remark}

\subsection{Complexity Analysis}
To further observe the performance of the U-B-MDS-MB-XTSPIR scheme, its complexity is analysed in this subsection.

\subsubsection*{Query Complexity}
For the queries of each user $m$ \eqref{query:1}, the user evaluates $F_m\lambda$ polynomials of degree less than $N$ at $N$ points for $S=K$ rounds. Notice from \cite{Von} that
the evaluation of a $k$-th degree polynomial at $k+1$ arbitrary points can be done in ${\mathcal{O}}(k(\log k)^2\log\log k)$ arithmetic operations.
Thus, the queries of user $m$ achieve a complexity at most $\mathcal{O}(KF_m\lambda N(\log N)^2\log\log N)$.

\subsubsection*{Server Computation Complexity}
For server response \eqref{answers}, each server $n$ first computes the product
of the $M+2$ elements $\phi_j^{s}(\alpha_n),Q_{j}^{(f_m),m,s}(\alpha_n),m\in[M],\varphi_j^{(f_1,\ldots,f_M)}(\alpha_n)$ for $F\lambda$ times, and then generates the response \eqref{answers} by taking the sum of the $F\lambda$ products and the stored randomness $\widetilde{\mathcal{Z}}_n^s$, where $F=F_1F_2\ldots F_M$.
Notably, the complexity of evaluating the intermediate polynomials $\phi_j^{s}(\alpha),j\in[\lambda]$ at point $\alpha=\alpha_n$ is negligible since the polynomials are constructed independently of data files and thus can be computed at servers a priori during off-peak hours to reduce the latency of server computation. Hence, computing the responses \eqref{answers} achieve the complexity $\mathcal{O}(KMF\lambda)$ for $S=K$ rounds.

\subsubsection*{Decoding Complexity}
For decoding complexity, in each round, each user first decodes the answer polynomial $A^s(\alpha)$ from a RS codeword of dimension $N$ and then evaluates the polynomial at $\lambda<N$ points.
By \cite{Lin,Gao} and \cite{Von}, such operations of RS decoding and evaluations can be done within the complexity $\mathcal{O}(N(\log N)^2\log\log N)$.
Thus, decoding achieves a complexity at most $\mathcal{O}(KN(\log N)^2\log\log N)$ for $S=K$ rounds.

\section{Related Work and Comparison}\label{comparison}
The most valuable aspect of the proposed U-B-MDS-MB-XTSPIR scheme is that a new form of interference alignment to PIR is created, based on the structure inspired by Lagrange polynomials. To see the innovations of the scheme in perspective, let us compare our scheme with the U-B-MDS-TPIR scheme \cite{Tajeddine222}, the U-B-MDS-XTPIR scheme \cite{MDS-X-security} and the MB-XTSPIR scheme \cite{B-PIR} that are most relevant work to our, where U-B-MDS-MB-XTSPIR includes as special cases the settings of U-B-MDS-TPIR, obtained by setting $M=1,X=0$ and eliminating server privacy, U-B-MDS-XTSPIR, obtained by setting $M=1$ and eliminating server privacy, and MB-XTSPIR, obtained by setting $K=1,B=U=0$.

In general, to privately retrieve the desired file from server responses, all the coded schemes above employ the structure inspired by MDS codes to create data storage (satisfying MDS property or/and security constraint) and private queries, and the response of each server can be viewed as the inner products of MDS coded private query vectors and MDS coded (secure) stored data vectors.
Notably, the MDS coded storage and queries are designed with the same code parameters $\boldsymbol{\alpha}=(\alpha_1,\alpha_2,\ldots,\alpha_N)$, such that the responses of all the servers constitute MDS codewords
because the Hadamard product of the two MDS codes employed by the data storage and queries with the same parameter $\boldsymbol{\alpha}$ is again an MDS code with the parameter $\boldsymbol{\alpha}$, which can efficiently resist the Byzantine servers and unresponsive servers because MDS codes have the maximum Hamming distance.


However, for different PIR settings, data storage and queries are created with different MDS coded structure.
For the settings of MDS-TPIR and U-B-MDS-TPIR, the schemes in \cite{MDS-TPIR,Tajeddine222} employ the Reed-Solomon (RS) coded structure to create storage and queries, such that the data in data storage and query design are encoded as coefficients of polynomial functions.
To extend PIR setting to include secure storage constraint, the idea of Cross Subspace Alignment (CSA) is introduced in \cite{X-security} and then was generalized to the settings of U-B-MDS-XTPIR \cite{MDS-X-security} and MB-XTSPIR \cite{B-PIR}. CSA employs a Cauchy-Vandermonde MDS coded structure to construct data storage and queries, which creates a form of interference alignment to separate the desired terms and interference in server responses,
such that the desired terms appear along the dimensions corresponding to the Cauchy part and interference is aligned as much as possible along the dimensions corresponding to the Vandermonde part.
Nonetheless, the current structure of coded storage and queries (for examples, RS coded structure in \cite{MDS-TPIR,Tajeddine222} and Cauchy-Vandermonde structure of CSA codes \cite{X-security,MDS-X-security,B-PIR}) are not directly useful for the general setting of U-B-MDS-MB-XTSPIR considered in this paper, which will be explained in the following subsections through some simple examples.
That is, to efficiently retrieve the desired file for the U-B-MDS-MB-XTSPIR problem, it may be necessary to develop a new form of interference alignment.
To this end, our codes use the structure of Lagrange interpolation polynomials to create data storage and queries, such that the user can interpolate some polynomials from the server responses and then evaluate the polynomials to obtain desired symbols.

Next, we will use some simple examples to illustrate the intuitive comparisons above.
Before that, we outline a general designed framework of U-B-MDSXTSPIR, which can simplify the scheme description and make clearer comparison.
Furthermore, for convenience, we leave out the constraint of server privacy/blind privacy, i.e., set the noise polynomial $\psi^s(\alpha)=0$ in the following subsections.

\subsection{General Framework of U-B-MDS-MB-XTSPIR}
When server privacy and blind privacy are not considered, the key of our U-B-MDS-MB-XTSPIR scheme in each round $s\in[S]$ is to construct a secret-shared storage polynomial $\varphi_i^{(f_1,\ldots,f_M)}(\alpha)$ to securely encode
the $i$-th row data of the file $\mathbf{W}^{(f_1,\ldots,f_M)}$ for any $i\in[\lambda]$ and $f_1\in[F_1],\ldots,f_M\in[F_M]$, a secret-shared query polynomial $Q_i^{(f_m),m,s}(\alpha)$ for any $m\in[M]$ and $f_m\in[F_m]$, and an intermediate polynomial $\phi_i^{s}(\alpha)$ for any $i\in[\lambda]$, such that each user can recover an answer polynomial $A^s(\alpha)$ from the answers received from $N$ servers and then decodes $P=N-(K+X+T_1+\ldots+T_M+2B+U-1)$ desired symbols, where the answer polynomial $A^s(\alpha)$ is given by
\begin{IEEEeqnarray}{l}\label{answers:example}
A^s(\alpha)=\sum\limits_{f_1\in[F_1],\ldots,f_M\in[F_M]}\sum\limits_{i\in[\lambda]}\phi_i^{s}(\alpha)\notag\\
\quad\quad\quad\times\bigg(\prod\limits_{m\in[M]}Q_{i}^{(f_m),m,s}(\alpha)\bigg)\cdot\varphi_i^{(f_1,\ldots,f_M)}(\alpha).
\end{IEEEeqnarray}

Then, the data stored at server $n$ and the query sent by user $m$ to server $n$ are given by evaluating the storage polynomials $\varphi_i^{(f_1,\ldots,f_M)}(\alpha)$ and the query polynomials $Q_i^{(f_m),m,s}(\alpha)$ at point $\alpha=\alpha_n$, as shown in \eqref{stored data} and \eqref{query:1}, respectively.
Upon receiving the queries from the $M$ users, server $n$ responds with the answer $A^s(\alpha_n)$, i.e., the evaluation of $A^s(\alpha)$ at $\alpha=\alpha_n$, as shown in \eqref{answers}.
Finally, each user recovers the polynomial $A^{s}(\alpha)$ from the $N$ server answers $A^s(\alpha_1),\ldots,A^s(\alpha_N)$ and then obtains $P$ desired symbols, as shown in \eqref{evaluating:1}.

Based on this, we can design an achievable scheme after giving storage polynomials, query polynomials and  intermediate polynomials satisfying decoding constraint.
Notably, the designed framework contains the U-B-MDS-TPIR scheme \cite{Tajeddine222}, the U-B-MDS-XTPIR scheme \cite{MDS-X-security} and the MB-XTSPIR scheme \cite{B-PIR} as special cases.
Thus, when we compare these schemes in the following examples, it is enough to just illustrate the design details of storage polynomials, query polynomials and  intermediate polynomials.

For a clearer comparison, in the following examples, we fix the parameters $K=2$ in the case of MDS coded storage (or $K=1$ in the case of replication storage), $X=2$ in the case of secure storage constraint (or $X=0$, without secure storage constraint), $\lambda=3,T_1=\ldots=T_M=2,B=0,U=0$, and freely choose $N$ such that $P=\lambda=N-(K+X+T_1+\ldots+T_M+2B+U-1)=3$.

\subsection{Illustrative Example for U-B-MDS-TPIR}\label{example: U-B-MDS-TPIR scheme}
In this subsection, we illustrate the U-B-MDS-TPIR scheme \cite{Tajeddine222} by describing the coded structure of its storage polynomials and query polynomials, where all the intermediate polynomials are set to be $1$ in the scheme. U-B-MDS-MB-XTSPIR includes U-B-MDS-TPIR as a special case by setting $M=1,X=0$ and eliminating server privacy, and thus we choose $N=6$, where the queries consist of $S=K=2$ rounds for retrieving the $\theta_1$-th file.

Each file $\mathbf{W}^{(f_1)}$ for any $f_1\in[F_1]$ is the form of
\begin{IEEEeqnarray}{rCl}\label{exam:file symbols2}
\mathbf{W}^{(f_1)} =\left[
  \begin{array}{cc}
    w^{(f_1)}_{1,1} & w^{(f_1)}_{1,2}  \\
    w^{(f_1)}_{2,1} & w^{(f_1)}_{2,2}\\
    w^{(f_1)}_{3,1} & w^{(f_1)}_{3,2}\\
\end{array}
\right].
\end{IEEEeqnarray}

For every $f_1\in[F_1]$ and $i\in[3]$, the scheme creates a RS coded storage polynomial as
\begin{IEEEeqnarray}{c}\label{storage:camilla}
\varphi_i^{(f_1)}(\alpha)=w_{i,1}^{(f_1)}+w_{i,2}^{(f_1)}\alpha.
\end{IEEEeqnarray}

During rounds $s=1,2$, the RS coded secret-shared query polynomials are constructed as
\begin{IEEEeqnarray}{l}
Q^{(f_1),1,1}_{i}(\alpha)=z_{i,1}^{(f_1),1,1}+z_{i,2}^{(f_1),1,1}\alpha \notag\\
\quad\quad\quad\quad\quad\quad\quad\quad+\left\{
\begin{array}{@{}ll}
\alpha^4, &\mathrm{if}\,\,f_1=\theta_1,i=1\\
\alpha^2, &\mathrm{if}\,\,f_1=\theta_1,i=2\\
0, &\mathrm{otherwise}
\end{array}
\right. \label{queri:camilla1}
\end{IEEEeqnarray}
and
\begin{IEEEeqnarray}{l}
Q^{(f_1),1,2}_{i}(\alpha)=z_{i,1}^{(f_1),1,2}+z_{i,2}^{(f_1),1,2}\alpha \notag\\
\quad\quad\quad\quad\quad\quad\quad\quad+\left\{
\begin{array}{@{}ll}
\alpha^5, &\mathrm{if}\,\, f_1=\theta_1,i=2\\
\alpha^3, &\mathrm{if}\,\,f_1=\theta_1,i=3\\
0, &\mathrm{if}\,\, \mathrm{otherwise}
\end{array}
\right.,\label{queri:camilla2}
\end{IEEEeqnarray}
respectively, where $z_{i,1}^{(f_1),1,s}$ and $z_{i,2}^{(f_1),1,s}$ are random noises that are used to guarantee $T_1=2$-private queries, and the coefficients $\alpha^4,\alpha^2$ in $Q^{(f_1),1,1}_{i}(\alpha)$ and $\alpha^5,\alpha^3$ in $Q^{(f_1),1,2}_{i}(\alpha)$ are used to separate the desired terms from the interference. This can be checked by expanding the answer polynomial $A^s(\alpha)$ \eqref{answers:example}, as follows. In round $s=1$,
\begin{IEEEeqnarray}{l}
A^1(\alpha)=\sum\limits_{f_1\in[F_1]}\sum\limits_{i\in[3]}Q^{(f_1),1,1}_{i}(\alpha)\cdot\varphi_i^{(f_1)}(\alpha) \notag \\
=\underbrace{I_0^1+I_1^1\alpha+I_2^1\alpha^2}_{\text{Interference Alignment}}+\underbrace{w_{2,2}^{(\theta_1)}\alpha^3+w_{1,1}^{(\theta_1)}\alpha^4+w_{1,2}^{(\theta_1)}\alpha^5}_{\text{Desired Terms}}, \label{example:111} \IEEEeqnarraynumspace
\end{IEEEeqnarray}
where the desired symbols $w_{2,2}^{(\theta_1)},w_{1,1}^{(\theta_1)},w_{1,2}^{(\theta_1)}$ appear along the terms $\alpha^3,\alpha^4,\alpha^5$, respectively, and the interference $I_0^1,I_1^1,I_2^1$ are aligned within a $3$ dimensional space. Thus, the user can recover the desired symbols $w_{2,2}^{(\theta_1)},w_{1,1}^{(\theta_1)},w_{1,2}^{(\theta_1)}$ by interpolating $A^1(\alpha)$ from the $N=6$ server answers in round $s=1$.

Similarly, the answer polynomial in round $s=2$ can be expanded as
\begin{IEEEeqnarray}{l}
A^2(\alpha)=\underbrace{I_0^2+I_1^2\alpha+I_2^2\alpha^2}_{\text{Interference Alignment}} \notag\\
+\underbrace{w_{3,1}^{(\theta_1)}\alpha^3+w_{3,2}^{(\theta_1)}\alpha^4+w_{2,1}^{(\theta_1)}\alpha^5}_{\text{Desired Terms}}+\underbrace{w_{2,2}^{(\theta_1)}\alpha^6}_{\text{Known Interference}}. \label{example:222}
\end{IEEEeqnarray}
Apparently, the answer polynomial $A^2(\alpha)$ occupies a $7$ dimensional space. The user cannot recover $A^2(\alpha)$ since it just receives $N=6$ dimensional answers.
However, the symbol $w_{2,2}^{(\theta_1)}$  has been recovered and thus the user can eliminate the interference from the symbol $w_{2,2}^{(\theta_1)}$ in the answers, and then decodes the desired symbols $w_{3,1}^{(\theta_1)},w_{3,2}^{(\theta_1)},w_{3,3}^{(\theta_1)}$.

\begin{Remark}\label{chuanxing}
It is straightforward to observe that, when the answers in rounds $s=1,2$ are received, the user must first decode the desired symbols $w_{2,2}^{(\theta_1)},w_{1,1}^{(\theta_1)},w_{1,2}^{(\theta_1)}$ from the answers in round $s=1$. Then, after eliminating the interference from the previously retrieved desired symbols in the answers in round $s=2$, the user can further decode the symbols $w_{3,1}^{(\theta_1)},w_{3,2}^{(\theta_1)},w_{3,3}^{(\theta_1)}$. That is, the user must decode desired symbols \emph{serially} in the order of rounds.
\end{Remark}

\subsection{Illustrative Example for U-B-MDS-XTPIR}\label{example:U-B-MDS-XTPIR}
In this subsection, we describe the U-B-MDS-XTPIR scheme \cite{MDS-X-security} by presenting the coded structure of its storage polynomials, query polynomials and intermediate polynomials. U-B-MDS-MB-XTSPIR includes U-B-MDS-XTPIR as a special case by setting $M=1$ and eliminating server privacy, and thus we choose $N=8$, where the queries consist of $S=K=2$ rounds for retrieving the $\theta_1$-th file.



Similar to \eqref{exam:file symbols2}, the file $\mathbf{W}^{(f_1)}$ for any $f_1\in[F_1]$ is
\begin{IEEEeqnarray}{rCl}\label{exam:file symbols123}
\mathbf{W}^{(f_1)} =\left[
  \begin{array}{cc}
    w^{(f_1)}_{1,1} & w^{(f_1)}_{1,2}  \\
    w^{(f_1)}_{2,1} & w^{(f_1)}_{2,2}\\
    w^{(f_1)}_{3,1} & w^{(f_1)}_{3,2}\\
\end{array}
\right].
\end{IEEEeqnarray}

Notably, when $X$-secure constraint is considered, it is not an efficient solution to directly extend the U-B-MDS-TPIR scheme in Section \ref{example: U-B-MDS-TPIR scheme} to retrieve the desired file $\mathbf{W}^{(\theta_1)}$ \eqref{exam:file symbols123}.
This is because, the RS coded storage polynomial \eqref{storage:camilla} in U-B-MDS-TPIR scheme employs the identical structure ($1,\alpha,\alpha^2,\ldots$) to encode each raw data of all files, which forces the user to retrieve all the coefficients of the (secure) storage polynomials $\varphi_i^{(\theta_1)}(\alpha),i\in[3]$ for recovering the desired file, see \eqref{example:111} and \eqref{example:222}.
Under $X=2$-secure constraint, following the RS coded storage structure of U-B-MDS-TPIR scheme, the secret-shared storage polynomial $\varphi_i^{(\theta)}(\alpha)$ is given by
\begin{IEEEeqnarray}{c}\notag
\varphi_i^{(\theta_1)}(\alpha)=w_{i,1}^{(\theta_1)}+w_{i,2}^{(\theta_1)}\alpha+z_{i,3}^{(\theta_1)}\alpha^2+z_{i,4}^{(\theta_1)}\alpha^3.
\end{IEEEeqnarray}
To retrieve the desired file, U-B-MDS-TPIR scheme will decode some redundant noises $z_{i,3}^{(\theta_1)},z_{i,4}^{(\theta_1)}$, which results in low retrieval rate.

To solve this problem, the idea of CSA is introduced, which employs distinct MDS coded storage structure for all data rows of each file, but the identical structure for the same data row of all the files, where the goal of the former is to distinguish all the symbols of each file and the latter is to create interference alignment opportunities over all the files.

Based on this idea, the MDS coded secret-shared storage polynomial $\varphi_i^{(f_1)}(\alpha)$ for any $f_1\in[F_1]$ and $i\in[3]$ is given by
\begin{IEEEeqnarray}{l}\label{example:CSA1}
\varphi_i^{(f_1)}(\alpha)=w_{i,1}^{(f_1)}+w_{i,2}^{(f_1)}(f_i-\alpha)\notag\\
\quad\quad\quad\quad\quad+z_{i,3}^{(f_1)}(f_i-\alpha)^2+z_{i,4}^{(f_1)}(f_i-\alpha)^3,
\end{IEEEeqnarray}
where $f_1,f_2,f_3\in\mathbb{F}_q$ are distinct from $\alpha_1,\ldots,\alpha_N$.
The MDS coded secret-shared query polynomials in rounds $s=1,2$ are
\begin{IEEEeqnarray}{l}
Q^{(f_1),1,1}_{i}(\alpha)=z_{i,1}^{(f_1),1,1}(f_i-\alpha)+z_{i,2}^{(f_1),1,1}(f_i-\alpha)^2 \notag\\
\quad\quad\quad\quad\quad\quad\quad\quad\quad\quad+\left\{
\begin{array}{@{}ll}
1, &\mathrm{if}\,\,f_1=\theta_1\\
0, &\mathrm{otherwise}
\end{array}
\right.. \label{exam:CSA:1234}
\end{IEEEeqnarray}
and
\begin{IEEEeqnarray}{l}
Q^{(f_1),1,2}_{i}(\alpha)=z_{i,1}^{(f_1),1,2}(f_i-\alpha)^2+z_{i,2}^{(f_1),1,2}(f_i-\alpha)^3 \notag\\
\quad\quad\quad\quad\quad\quad\quad\quad\quad\quad+\left\{
\begin{array}{@{}ll}
1, &\mathrm{if}\,\,f_1=\theta_1\\
0, &\mathrm{otherwise}
\end{array}
\right.. \label{exam:CSA:12341234}
\end{IEEEeqnarray}

The intermediate polynomial $\phi_i^{s}(\alpha)$ is given by
\begin{IEEEeqnarray}{c}\label{intermediate polys}
\phi_i^{s}(\alpha)=\frac{1}{(f_i-\alpha)^s},\quad\forall\, i\in[3].
\end{IEEEeqnarray}

The interference alignment rule of CSA idea can be checked by extending  the answer polynomial $A^s(\alpha)$ \eqref{answers:example}.  In round $s=1$,
\begin{IEEEeqnarray}{rCl}
A^1(\alpha)&=&\sum\limits_{f_1\in[F_1]}\sum\limits_{i\in[3]}\phi_i^{1}(\alpha)\cdot Q^{(f_1),1,1}_{i}(\alpha)\cdot\varphi_i^{(f_1)}(\alpha) \notag \\
&=&\underbrace{\frac{1}{f_1-\alpha}w_{1,1}^{(\theta_1)}+\frac{1}{f_2-\alpha}w_{2,1}^{(\theta_1)}+\frac{1}{f_3-\alpha}w_{3,1}^{(\theta_1)}}_{\text{Desired Terms}}\notag\\
&&+\underbrace{I_0^1+I_1^1\alpha+I_2^1\alpha^2+I_3^1\alpha^3+I_4^1\alpha^4}_{\text{Interference Alignment}}. \label{example:align}
\end{IEEEeqnarray}
The desired symbols $w_{1,1}^{(\theta_1)},w_{2,1}^{(\theta_1)},w_{3,1}^{(\theta_1)}$ appear along the Cauchy terms $\frac{1}{f_1-\alpha},\frac{1}{f_2-\alpha},\frac{1}{f_3-\alpha}$, respectively, and the interference $I_0^1,\ldots,I_4^1$ are aligned within a $5$ dimensional space.
Then, the user can recover $A^1(\alpha)$ from the $N=8$ dimensional answers, and obtains the desired symbols $w_{1,1}^{(\theta_1)},w_{2,1}^{(\theta_1)},w_{3,1}^{(\theta_1)}$ in round $s=1$.

Similarly, the answer polynomial in round $s=2$ can be denoted by
\begin{IEEEeqnarray}{rCl}
A^2(\alpha)&=&\underbrace{\frac{1}{(f_1-\alpha)^2}w_{1,1}^{(\theta_1)}+\frac{1}{(f_2-\alpha)^2}w_{2,1}^{(\theta_1)}+\frac{1}{(f_3-\alpha)^2}w_{3,1}^{(\theta_1)}}_{\text{Known Interference}}\notag\\
&&+\underbrace{\frac{1}{f_1-\alpha}w_{1,2}^{(\theta_1)}+\frac{1}{f_2-\alpha}w_{2,2}^{(\theta_1)}+\frac{1}{f_3-\alpha}w_{3,2}^{(\theta_1)}}_{\text{Desired Terms}}\notag\\
&&+\underbrace{I_0^2+I_1^2\alpha+I_2^2\alpha^2+I_3^2\alpha^3+I_4^2\alpha^4}_{\text{Interference Alignment}}. \notag
\end{IEEEeqnarray}
The user first eliminates the interference from the desired symbols $w_{1,1}^{(\theta_1)},w_{2,1}^{(\theta_1)},w_{3,1}^{(\theta_1)}$, and then recovers $w_{1,2}^{(\theta_1)},w_{2,2}^{(\theta_1)},w_{3,2}^{(\theta_1)}$ from the $N=8$ dimensional answers in round $s=2$.

\begin{Remark}\label{chuanxing2}
Similar to Remark \ref{chuanxing}, when the answers in rounds $s=1,2$ are received, the user must decode desired symbols \emph{serially} in the order of rounds.
\end{Remark}

\subsection{Illustrative Example for MB-XTSPIR}\label{example:MB}
In this subsection, we describe the MB-XTSPIR scheme \cite{B-PIR} by presenting the coded structure of its storage polynomials, query polynomials and intermediate polynomials. U-B-MDS-MB-XTSPIR includes MB-XTSPIR as a special case by setting $K=1,B=U=0$. We set $M=2$ and thus choose $N=9$, where the scheme consists of $S=1$ round.

Here, each file $\mathbf{W}^{(f_1,f_2)}$ for any $f_1\in[F_1],f_2\in[F_2]$ is the form of
\begin{IEEEeqnarray}{rCl}\label{exam:file symbols111}
\mathbf{W}^{(f_1,f_2)} =\left[
  \begin{array}{c}
    w^{(f_1,f_2)}_{1,1} \\
    w^{(f_1,f_2)}_{2,1} \\
    w^{(f_1,f_2)}_{3,1} \\
\end{array}
\right].
\end{IEEEeqnarray}


In essence, in the MB-XTSPIR scheme \cite{B-PIR},  the coded structure of storage polynomials, query polynomials and intermediate polynomials are similar to the ones in round $s=1$ of U-B-MDS-XTPIR scheme.
Specifically, the secret-shared storage polynomial $\varphi_i^{(f_1,f_2)}(\alpha)$ for every $f_1\in[F_1],f_2\in[F_2]$ and $i\in[3]$ is constructed similar to \eqref{example:CSA1}, given by
\begin{IEEEeqnarray}{c}\notag
\varphi_i^{(f_1,f_2)}(\alpha)=w_{i,1}^{(f_1,f_2)}+z_{i,2}^{(f_1,f_2)}(f_i-\alpha)+z_{i,3}^{(f_1,f_2)}(f_i-\alpha)^2.
\end{IEEEeqnarray}
The secret-shared query polynomial for each user $m=1,2$ follows the similar structure to \eqref{exam:CSA:1234},  given by
\begin{IEEEeqnarray}{l}
Q^{(f_m),m,1}_{i}(\alpha)=z_{i,1}^{(f_m),m,1}(f_i-\alpha)+z_{i,2}^{(f_m),m,1}(f_i-\alpha)^2 \notag\\
\quad\quad\quad\quad\quad\quad\quad\quad\quad\quad\quad+\left\{
\begin{array}{@{}ll}
1, &\mathrm{if}\,\,f_m=\theta_m\\
0, &\mathrm{otherwise}
\end{array}
\right..
\end{IEEEeqnarray}
The intermediate polynomial $\phi_i^{1}(\alpha)$ is the same as \eqref{intermediate polys}, i.e.,
\begin{IEEEeqnarray}{c}\notag
\phi_i^{1}(\alpha)=\frac{1}{f_i-\alpha},\quad\forall\, i\in[3].
\end{IEEEeqnarray}

Then, similar to \eqref{example:align}, the answer polynomial $A^1(\alpha)$ \eqref{answers:example} has
\begin{IEEEeqnarray}{rCl}
A^1(\alpha)&=&\sum\limits_{f_1\in[F_1],f_2\in[F_2]}\sum\limits_{i\in[3]}\phi_i^{1}(\alpha)\notag\\
&&\quad\quad\quad\cdot Q_{i}^{(f_1),1,1}(\alpha)\cdot Q_{i}^{(f_2),2,1}(\alpha)\cdot\varphi_i^{(f_1,f_2)}(\alpha) \notag\\
&=&\underbrace{\frac{1}{f_1-\alpha}w_{1,1}^{(\theta_1,\theta_2)}+\frac{1}{f_2-\alpha}w_{2,1}^{(\theta_1,\theta_2)}+\frac{1}{f_3-\alpha}w_{3,1}^{(\theta_1,\theta_2)}}_{\text{Desired Terms}}\notag\\
&&+\underbrace{I_0+I_1\alpha+I_2\alpha^2+I_3\alpha^3+I_4\alpha^4+I_5\alpha^5}_{\text{Interference Alignment}}. \label{MDS-CSA:1} \IEEEeqnarraynumspace
\end{IEEEeqnarray}
Each user can recover the desired symbols $w_{1,1}^{(\theta_1,\theta_2)},w_{2,1}^{(\theta_1,\theta_2)},w_{3,1}^{(\theta_1,\theta_2)}$ from the $N=9$ dimensional answers.

\subsection{Illustrative Example for MB-MDS-XTSPIR}
In this subsection, we consider the setting of MB-XTSPIR with MDS coded storage (MB-MDS-XTSPIR).
Similar to Section \ref{example:MB}, we try to generalize the U-B-MDS-XTPIR scheme in Section \ref{example:U-B-MDS-XTPIR} to the setting of MB-MDS-XTSPIR.
 We will set $M=2$, and thus choose $N=10$, where the scheme consists of $S=K=2$ round.

Each file $\mathbf{W}^{(f_1,f_2)}$ for any $f_1\in[F_1],f_2\in[F_2]$ is the form of
\begin{IEEEeqnarray}{rCl}\label{exam:file symbols222}
\mathbf{W}^{(f_1,f_2)} =\left[
  \begin{array}{cc}
    w^{(f_1,f_2)}_{1,1} & w^{(f_1,f_2)}_{1,2}  \\
    w^{(f_1,f_2)}_{2,1} & w^{(f_1,f_2)}_{2,2}\\
    w^{(f_1,f_2)}_{3,1} & w^{(f_1,f_2)}_{3,2}\\
\end{array}
\right].
\end{IEEEeqnarray}

Similar to \eqref{example:CSA1}, for every $f_1\in[F_1],f_2\in[F_2]$ and $i\in[3]$, the secret-shared storage polynomial $\varphi_i^{(f_1,f_2)}(\alpha)$ is given by
\begin{IEEEeqnarray}{l}\notag
\varphi_i^{(f_1,f_2)}(\alpha)=w_{i,1}^{(f_1,f_2)}+w_{i,2}^{(f_1,f_2)}(f_i-\alpha)\notag\\
\quad\quad\quad\quad\quad+z_{i,2}^{(f_1,f_2)}(f_i-\alpha)^2+z_{i,3}^{(f_1,f_2)}(f_i-\alpha)^3.
\end{IEEEeqnarray}

Similar to \eqref{exam:CSA:1234} and \eqref{exam:CSA:12341234}, the secret-shared query polynomials of user $m$ in rounds $s=1,2$ are given by
\begin{IEEEeqnarray}{l}
Q^{(f_m),m,1}_{i}(\alpha)=z_{i,1}^{(f_m),m,1}(f_i-\alpha)+z_{i,2}^{(f_m),m,1}(f_i-\alpha)^2 \notag\\
\quad\quad\quad\quad\quad\quad\quad\quad+\left\{
\begin{array}{@{}ll}
1, &\mathrm{if}\,\,f_m=\theta_m\\
0, &\mathrm{otherwise}
\end{array}
\right. \label{exam:CSA:1}
\end{IEEEeqnarray}
and
\begin{IEEEeqnarray}{l}
Q^{(f_m),m,2}_{i}(\alpha)=z_{i,1}^{(f_m),m,2}(f_i-\alpha)^2+z_{i,2}^{(f_m),m,2}(f_i-\alpha)^3 \notag\\
\quad\quad\quad\quad\quad\quad\quad\quad+\left\{
\begin{array}{@{}ll}
1, &\mathrm{if}\,\,f_m=\theta_m\\
0, &\mathrm{otherwise}
\end{array}
\right..
\end{IEEEeqnarray}


By \eqref{intermediate polys}, the intermediate polynomial $\phi_i^{s}(\alpha)$ is given by
\begin{IEEEeqnarray}{c}
\phi_i^{s}(\alpha)=\frac{1}{(f_i-\alpha)^s},\quad\forall\, i\in[3].
\end{IEEEeqnarray}

Similar to \eqref{example:align} and \eqref{MDS-CSA:1}, in round $s=1$, each user can decode the desired symbols $w^{(\theta_1,\theta_2)}_{1,1},w^{(\theta_1,\theta_2)}_{2,1},w^{(\theta_1,\theta_2)}_{3,1}$.
However, during round $s=2$, each user cannot recover the answer polynomial $A^2(\alpha)$ from the received answers, which can be checked by expanding $A^2(\alpha)$ as
\begin{IEEEeqnarray}{l}
A^2(\alpha)=\sum\limits_{f_1\in[F_1],f_2\in[F_2]}\sum\limits_{i\in[3]}\phi_i^{2}(\alpha)\notag\\
\quad\quad\quad\quad\quad\quad\cdot Q_{i}^{(f_1),1,2}(\alpha)\cdot Q_{i}^{(f_2),2,2}(\alpha)\cdot\varphi_i^{(f_1,f_2)}(\alpha) \notag\\
=\underbrace{\sum\limits_{i\in[3]}\frac{1}{(f_i-\alpha)^2}w_{i,1}^{(\theta_1,\theta_2)}}_{\text{Known Interference}}+\underbrace{\sum\limits_{i\in[3]}\frac{1}{f_i-\alpha}w_{i,2}^{(\theta_1,\theta_2)}}_{\text{Desired Terms}}\notag\\
+\underbrace{I_0^2+I_1^2\alpha+I_2^2\alpha^2+I_3^2\alpha^3+I_4^2\alpha^4+I_5^2\alpha^5+I_6^2\alpha^6+I_7^2\alpha^7}_{\text{Interference Alignment}}. \notag
\end{IEEEeqnarray}
Notably, each user only has an $N=10$ dimensional answers. However, even if the user eliminates the interference from the previously retrieved symbols $w^{(\theta_1,\theta_2)}_{1,1},w^{(\theta_1,\theta_2)}_{2,1},w^{(\theta_1,\theta_2)}_{3,1}$, and the remaining terms in $A^2(\alpha)$ occupy a $11$ dimensional space, which exceeds the dimensions of answers and means that the user cannot decode the symbols $w^{(\theta_1,\theta_2)}_{1,2},w^{(\theta_1,\theta_2)}_{2,2},w^{(\theta_1,\theta_2)}_{3,2}$ from the answers.
The reasons are explained as follows. Among the $10$ dimensional answer space, there are $P=3$ dimensions that are used to retrieve the desired symbols, and all the interference should be aligned within the remaining $7$ dimensional space. However, during round $s=2$, the scheme creates the query polynomial $Q^{(f_m),m,2}_{i}(\alpha)$ of degree $3$ for each user $m=1,2$, the intermediate polynomial $\phi_i^{2}(\alpha)=\frac{1}{(f_i-\alpha)^2}$ and the storage polynomial $\varphi_i^{(f_1,f_2)}(\alpha)$ of degree $3$, such that the interference are aligned within a $\deg(Q^{(f_1),1,2}_{i}(\alpha)\cdot Q^{(f_2),2,2}_{i}(\alpha)\cdot \phi_i^{2}(\alpha)\cdot\varphi_i^{(f_1,f_2)}(\alpha))+1=8$ dimensional space.

To solve the problem, during each round $s=1,2$, we design a Lagrange storage polynomial $\varphi_i^{(f_1,f_2)}(\alpha)$ of degree $3$ such that
\begin{IEEEeqnarray}{rClrCl}
\varphi_i^{(f_1,f_2)}(\beta_{i,1})&=&w_{i,1}^{(f_1,f_2)},\quad
\varphi_i^{(f_1,f_2)}(\beta_{i,2})&=&w_{i,2}^{(f_1,f_2)},\IEEEeqnarraynumspace\label{example:storage:1111}\\
\varphi_i^{(f_1,f_2)}(\beta_{i,3})&=&z_{i,3}^{(f_1,f_2)},\quad
\varphi_i^{(f_1,f_2)}(\beta_{i,4})&=&z_{i,4}^{(f_1,f_2)}, \IEEEeqnarraynumspace \label{example:storage:2222123}
\end{IEEEeqnarray}
a private Lagrange polynomial $Q^{(f_m),m,s}_{i}(\alpha)$ of degree $2$ for each user $m=1,2$ such that
\begin{IEEEeqnarray}{rCl}
Q^{(f_m),m,s}_{i}(\beta_{i,s})&=&\left\{
\begin{array}{@{}ll}
1, &\mathrm{if}\,\,f_m=\theta_m\\
0, &\mathrm{if}\,\, f_m\neq\theta_m
\end{array}
\right., \label{example:storage:11111} \\
Q^{(f_m),m,s}_{i}(\alpha_1)&=&z_{i,1}^{(f_m),m,s},\label{example:storage:11112} \\
Q^{(f_m),m,s}_{i}(\alpha_2)&=&z_{i,2}^{(f_m),m,s},\label{example:storage:11113}
\end{IEEEeqnarray}
and the intermediate polynomials $\phi_i^s(\alpha)$ of degree $2$ such that
\begin{IEEEeqnarray}{rClrClrCl}
\phi_1^s(\beta_{1,s})&=&1, \quad \phi_1^s(\beta_{2,s})&=&0, \quad \phi_1^s(\beta_{3,s})&=&0, \label{example:storage:222222} \\
\phi_2^s(\beta_{1,s})&=&0, \quad \phi_2^s(\beta_{2,s})&=&1, \quad \phi_2^s(\beta_{3,s})&=&0,  \\
\phi_3^s(\beta_{1,s})&=&0, \quad \phi_3^s(\beta_{2,s})&=&0, \quad \phi_3^s(\beta_{3,s})&=&1. \label{example:storage:2222}
\end{IEEEeqnarray}
In each round $s=1,2$, such coded structure based on Lagrange polynomials align $P=3$ desired symbols and the interference within a $\deg(Q^{(f_1),1,s}_{i}(\alpha)\cdot Q^{(f_2),2,s}_{i}(\alpha)\cdot \phi_i^{s}(\alpha)\cdot\varphi_i^{(f_1,f_2)}(\alpha))+1=10$ dimensional space, which exactly matches the $N=10$ dimensional answer space, i.e., the desired terms occupy $3$ dimensions and the interference are aligned within the remaining $7$ dimensions.

Then each user can recover the answer polynomial
\begin{IEEEeqnarray}{l}
A^s(\alpha)=\sum\limits_{f_1\in[F_1],f_2\in[F_2]}\sum\limits_{i\in[3]}\phi_i^s(\alpha)\cdot Q^{(f_1),1,s}_{i}(\alpha)\notag\\
\quad\quad\quad\quad\quad\quad\cdot Q^{(f_2),2,s}_{i}(\alpha)\cdot \varphi_i^{(f_1,f_2)}(\alpha).\notag 
\end{IEEEeqnarray}
Finally, by \eqref{example:storage:1111}-\eqref{example:storage:2222}, each user can evaluate $A^s(\alpha)$ at $\alpha=\beta_{1,s},\beta_{2,s},\beta_{3,s}$ and obtains $w_{1,s}^{(\theta_1,\theta_2)},w_{2,s}^{(\theta_1,\theta_2)},w_{3,s}^{(\theta_1,\theta_2)}$.


To conclude, the current structure of coded storage and queries (typically, RS coded structure in \cite{MDS-TPIR,Tajeddine222} and Cauchy-Vandermonde structure of CSA codes \cite{X-security,MDS-X-security,B-PIR}) are not directly generalized to the general setting of U-B-MDS-MB-XTSPIR considered in this paper.
To efficiently retrieve the desired file for U-B-MDS-MB-XTSPIR problem, we create a new form of interference alignment to PIR, based on the structure of Lagrange interpolation polynomials.

\begin{Remark}
Apparently, in our scheme based on Lagrange polynomials, the user can \emph{parallel} decode desired symbols from the answers in rounds $s=1,2$, which improves the efficiency of retrieving desired file, compared to \emph{serial} decoding in the order of rounds \cite{Tajeddine222,MDS-X-security} by Remarks \ref{chuanxing} and \ref{chuanxing2}.
\end{Remark}
\begin{Remark}
Lagrange coded computing (LCC) is initially introduced in \cite{LCC} for evaluating a multivariate polynomial over a batch of dataset, and then was generalized to solve the problems of private polynomial computing \cite{Raviv PPC} and distributed (secure) matrix multiplication \cite{EP code,MatDot code,Qian Yu,batch matrix}.

In Private Polynomial Computing (PPC), the user wishes to privately compute a polynomial function evaluations over the files. The PPC scheme in \cite{Raviv PPC} first resorts to the U-B-MDS-TPIR scheme in \cite{Tajeddine222} to retrieve all the coefficients of some composite polynomial functions, then recovers the composite polynomials, and finally evaluates the composite polynomials to obtain the desired evaluations. Actually, the scheme in \cite{Raviv PPC} can be viewed as a direct generalization of U-B-MDS-TPIR scheme in \cite{Tajeddine222}.

In distributed (secure) matrix multiplication, the user wishes to compute the multiplication task $\mathbf{A}\mathbf{B}$ of two massive matrices $\mathbf{A}$ and $\mathbf{B}$.
The schemes based on LCC \cite{EP code,MatDot code,Qian Yu,batch matrix} first convert the multiplication task to the problem of (securely) computing the pairwise products $(\mathbf{A}_1\mathbf{B}_1,\ldots,\mathbf{A}_L\mathbf{B}_L)$ of two batch of matrices $\mathbf{A}_1,\ldots,\mathbf{A}_L$ and $\mathbf{B}_1,\ldots,\mathbf{B}_L$ for some integer $L$, where $\mathbf{A}_1,\ldots,\mathbf{A}_L$ and $\mathbf{B}_1,\ldots,\mathbf{B}_L$ are the coded sub-matrices of $\mathbf{A}$ and $\mathbf{B}$, respectively.
Then, in encoding phase, the user creates two secret-shared Lagrange polynomials $\widetilde{\mathbf{A}}(\alpha),\widetilde{\mathbf{B}}(\alpha)$ to encode the two batch of matrices, respectively, such that
\begin{IEEEeqnarray}{l}
\widetilde{\mathbf{A}}(\beta_{\ell})=\left\{
\begin{array}{@{}ll}
\mathbf{A}_{\ell}, &\mathrm{if}\,\,\ell\in[L] \\
\mathbf{Z}_{\ell}^{\mathbf{A}}, &\mathrm{if}\,\,\ell\in[L+1:L+X] \\
\end{array}
\right.\label{MM:1}
\end{IEEEeqnarray}
and
\begin{IEEEeqnarray}{l}
\widetilde{\mathbf{B}}(\beta_{\ell})=\left\{
\begin{array}{@{}ll}
\mathbf{B}_{\ell}, &\mathrm{if}\,\,\ell\in[L] \\
\mathbf{Z}_{\ell}^{\mathbf{B}}, &\mathrm{if}\,\,\ell\in[L+1:L+X] \\
\end{array}
\right.,\label{MM:2}
\end{IEEEeqnarray}
where $\mathbf{Z}_{\ell}^{\mathbf{A}},\mathbf{Z}_{\ell}^{\mathbf{B}}$ are random noises.
In decoding phase, the user first recovers the product polynomial $\widetilde{\mathbf{A}}(\alpha)\widetilde{\mathbf{B}}(\alpha)$ and then evaluates it at $\alpha=\beta_1,\ldots,\beta_L$ to obtain the pairwise products $(\mathbf{A}_1\mathbf{B}_1,\ldots,\mathbf{A}_L\mathbf{B}_L)$.

In fact, our MB-MDS-XTSPIR scheme first employs the encoding idea \eqref{MM:1}-\eqref{MM:2} of LCC to create the secret-shared storage polynomials \eqref{example:storage:1111}-\eqref{example:storage:2222123}. Then by following decoding idea of LCC, we further employ Lagrange polynomials to create secret-shared query polynomials \eqref{example:storage:11111}-\eqref{example:storage:11113} and intermediate polynomials \eqref{example:storage:222222}-\eqref{example:storage:2222}, such that the user can interpolate some polynomials from server answers and then evaluates it to obtain desired symbols.
How to employ the structure inspired by Lagrange polynomials to create secret-shared queries under decoding constraint is the difficulty and our key innovation of applying the ideas of Lagrange coded computing to PIR problems.


\end{Remark}


\section{Conclusion}\label{conclusion}
In this paper, the problem of U-B-MDS-MB-XTSPIR was focused. By constructing an U-B-MDS-MB-XTSPIR scheme based on Lagrange encoding, we showed that the retrieval rate $1-\frac{K+X+T_1+\ldots+T_M+2B-1}{N-U}$
with secrecy rate $\frac{K+X+T_1+\ldots+T_M-1}{N-(K+X+T_1+\ldots+T_M+2B+U-1)}$ and finite field size $q\geq N+\max\{K, N-(K+X+T_1+\ldots+T_M+2B+U-1)\}$ is achievable for any number of files.

U-B-MDS-MB-XTSPIR generalizes the current optimal schemes as special cases of U-B-MDS-MB-XTSPIR including SPIR \cite{S-PIR1}, MDS-SPIR \cite{Wang-MDS-SPIR}, TSPIR \cite{wang-byzantine}, MDS-TSPIR \cite{wang-colluding-SPIR}, and MB-XTSPIR \cite{B-PIR}.
Further, when server privacy is not considered, the U-B-MDS-MB-XTSPIR scheme automatically yields the asymptotically optimal schemes for various special PIR cases \cite{Sun replicated,Ulukus_MDS,MDS-TPIR,Tajeddine222,MDS-X-security} as the number of files approaches infinity.
Thus, we conjecture that the general U-B-MDS-MB-XTSPIR scheme is also (asymptotically) optimal.
Naturally, this raises two promising open problems. One is to prove the optimality of the retrieval rate of the proposed solution, and the other is to characterize the minimal amount of randomness stored at servers for ensuring blind privacy and server privacy, which are valuable research directions for future work.

\begin{appendix}[Proof of Privacies]\label{proof:achevable}

In this appendix, we prove the privacies of the proposed U-B-MDS-MB-XTSPIR scheme.
\begin{Lemma}\label{lemma:achievability}
The proposed scheme in Section \ref{s-PPC scheme} is robust against $X$-secure data storage, user privacy, blind privacy and server privacy.
\end{Lemma}
\begin{IEEEproof}
It is sufficient to prove that the scheme satisfies the constraints \eqref{X security}, \eqref{Infor:priva cons}, and \eqref{Infor:priva cons2}.

\subsubsection*{User Privacy}
For any given $m\in[M]$, let $\mathcal{T}=\{n_1,\ldots,n_{T_m}\}\subseteq[N]$ be any $T_m$ of the $N$ server indices. By \eqref{symmetric:answer11} and \eqref{query:1}, the query elements $Q_{j}^{(f_m),m,s}(\alpha_{n_1}),$ $\ldots,Q_{j}^{(f_m),m,s}(\alpha_{n_{T_m}})$ sent to the servers $\mathcal{T}$ are protected by $T_m$ independent and uniform random noises for any $f_m\in[F_m],j\in[\lambda]$ and $s\in[S]$, as shown below.
\begin{IEEEeqnarray*}{l}
\left[
\begin{array}{@{}c@{}}
  Q_{j}^{(f_m),m,s}(\alpha_{n_1}) \\
  Q_{j}^{(f_m),m,s}(\alpha_{n_2}) \\
  \vdots \\
  Q_{j}^{(f_m),m,s}(\alpha_{n_{T_m}})
\end{array}
\right]=\underbrace{
\left[
\begin{array}{@{}c@{}}
  c_{j}^{(f_m),m,s}(\alpha_{n_1}) \\
  c_{j}^{(f_m),m,s}(\alpha_{n_2}) \\
  \vdots \\
  c_{j}^{(f_m),m,s}(\alpha_{n_{T_m}})
\end{array}
\right]}_{\triangleq\mathbf{c}_j^{(f_m),m,s}}\notag\\
\quad\quad\quad\quad\quad+\underbrace{
\left[
\begin{array}{@{}ccc@{}}
  h_1(\alpha_{n_1}) &  \ldots & h_{T_m}(\alpha_{n_1}) \\
  h_1(\alpha_{n_2}) & \ldots & h_{T_m}(\alpha_{n_2}) \\
  \vdots &  \ddots & \vdots \\
  h_1(\alpha_{n_{T_m}}) & \ldots & h_{T_m}(\alpha_{n_{T_m}}) \\
\end{array}
\right]}_{\triangleq{\mathbf{G}}_j^{s}}
\underbrace{
\left[
\begin{array}{@{}c@{}}
  z_{j,1}^{(f_m),m,s} \\
  \vdots \\
  z_{j,T_m}^{(f_m),m,s}
\end{array}
\right]}_{\triangleq\mathbf{z}_j^{(f_m),m,s}},\IEEEeqnarraynumspace\label{privacy:scheme1}
\end{IEEEeqnarray*}
where
\begin{IEEEeqnarray}{c}\label{constants}
c_{j}^{(f_m),m,s}(\alpha)=
\left\{\begin{array}{@{}ll}
\prod\limits_{v\in[T_m]}\frac{\alpha-\alpha_v}{\beta_{j,s}-\alpha_v}, &  \mathrm{if}\,\,f_m=\theta_m \\
0, &\mathrm{if}\,\,f_m\neq\theta_m
\end{array}
\right.
\end{IEEEeqnarray}
and
\begin{IEEEeqnarray}{c}\notag
h_\ell(\alpha)=\frac{\alpha-\beta_{j,s}}{\alpha_{\ell}-\beta_{j,s}}\cdot\prod\limits_{v\in[T_m]\backslash\{\ell\}}\frac{\alpha-\alpha_v}{\alpha_{\ell}-\alpha_v},\quad\forall\,\ell\in[T_m].
\end{IEEEeqnarray}

Note from P2-P4 that $\alpha_1,\ldots,\alpha_N,\beta_{j,s}$ are $N+1$ distinct elements for any $j\in[\lambda]$ and $s\in[S]$. Therefore, $\mathbf{G}_{j}^{s}$ is invertible over $\mathbb{F}_q$ by Lemma \ref{g-cauchy matrix}. Denote its inverse matrix by $(\mathbf{G}_{j}^{s})^{-1}$. Then, given any $f_m\in[F_m],j\in[\lambda]$ and $s\in[S]$,
\begin{IEEEeqnarray*}{rCl}
&&I\big(\{Q_{j}^{(f_m),m,s}(\alpha_{n})\}_{n\in\mathcal{T}};\theta_m\big)\\
&=&I\big(\mathbf{c}^{(f_m),m,s}_{j}+\mathbf{G}_{j}^{s}\cdot\mathbf{z}^{(f_m),m,s}_{j};\theta_m\big) \label{privacy:111} \\
&=&I\big((\mathbf{G}_{j}^{s})^{-1}\cdot\mathbf{c}^{(f_m),m,s}_{j}+\mathbf{z}^{(f_m),m,s}_{j};\theta_m\big) \\
&=&H\big((\mathbf{G}_{j}^{s})^{-1}\cdot\mathbf{c}^{(f_m),m,s}_{j}+\mathbf{z}^{(f_m),m,s}_{j}\big)\notag\\
&&\quad\quad\quad\quad-H\big((\mathbf{G}_{j}^{s})^{-1}\cdot\mathbf{c}^{(f_m),m,s}_{j}+\mathbf{z}^{(f_m),m,s}_{j}|\theta_m\big)\\
&\overset{(a)}{=}&H\big((\mathbf{G}_{j}^{s})^{-1}\cdot\mathbf{c}^{(f_m),m,s}_{j}+\mathbf{z}^{(f_m),m,s}_{j}\big)-H(\mathbf{z}^{(f_m),m,s}_{j}) \\
&\overset{(b)}{=}&0,  \label{privacy:222}
\end{IEEEeqnarray*}
where $(a)$ holds because $(\mathbf{G}_{j}^{s})^{-1}\cdot\mathbf{c}^{(f_m),m,s}_{j}$ is constant by \eqref{constants} when $\theta_m$ is given, and $\mathbf{z}^{(f_m),m,s}_{j}$ is generated independently of $\theta_m$, i.e., $H\big((\mathbf{G}_{j}^{s})^{-1}\cdot\mathbf{c}^{(f_m),m,s}_{j}+\mathbf{z}^{(f_m),m,s}_{j}|\theta_m\big)=H\big(\mathbf{z}^{(f_m),m,s}_{j}|\theta_m\big)=H(\mathbf{z}^{(f_m),m,s}_{j})$,
and $(b)$ follows from the fact that the elements $z_{j,1}^{(f_m),m,s},\ldots,z_{j,T_m}^{(f_m),m,s}$ in $\mathbf{z}^{(f_m),m,s}_{j}$ are i.i.d. uniformly over $\mathbb{F}_q$ and are independent of $(\mathbf{G}_{j}^{s})^{-1}\cdot\mathbf{c}^{(f_m),m,s}_{j}$, thus $\left(\mathbf{G}_{j}^{s}\right)^{-1}\cdot\mathbf{c}^{(f_m),m,s}_{j}+\mathbf{z}^{(f_m),m,s}_{j}$ and $\mathbf{z}^{(f_m),m,s}_{j}$ are identically and uniformly distributed over $\mathbb{F}_q^{T_m}$.

Then, for any $m\in[M]$,
\begin{IEEEeqnarray*}{rCl}
&&I({Q}_{\mathcal{T}}^{m};\theta_m)\notag\\
&\overset{(a)}{=}&I\big(\{Q_{j}^{(f_m),m,s}(\alpha_{n}):n\in\mathcal{T}\}_{j\in[\lambda], f_m\in[F_m],s\in[S]};\theta_m\big) \\
&\overset{(b)}{=}&\sum\limits_{s\in[S]}\sum\limits_{f_m\in[F_m]}\sum\limits_{j\in[\lambda]}I\big(\{Q_{j}^{(f_m),m,s}(\alpha_{n})\}_{n\in\mathcal{T}};\theta_m\big) \\
&=&0,
\end{IEEEeqnarray*}
where $(a)$ is due to \eqref{query:general} and \eqref{query:1}; $(b)$ follows from \eqref{symmetric:answer11} and the fact that the random noises $z_{j,1}^{(f_m),m,s},\ldots,z_{j,T_m}^{(f_m),m,s}$ that are used for protecting the queries $\{Q_{j}^{(f_m),m,s}(\alpha_{n})\}_{n\in\mathcal{T}}$ are independently and uniformly generated from $\mathbb{F}_q$ across all $j\in[\lambda],f_m\in[F_m]$ and $s\in[S]$. User-privacy follows by \eqref{Infor:priva cons}.

\subsubsection*{$X$-Security} Similar to user privacy, it is straight to prove $X$-security \eqref{X security} by Lemmas \ref{lemma:security} and \ref{g-cauchy matrix}.  

\subsubsection*{Blind Privacy and Server Privacy}
From \eqref{user randomness:round}, the private randomness at user $m$ is
\begin{IEEEeqnarray*}{c}
\mathcal{Z}_m=\{\mathcal{Z}_m^s\}_{s\in[S]},\quad\forall\,m\in[M].
\end{IEEEeqnarray*}
Moreover, for convenience, let $\Lambda^s(\alpha)$ be the first term of the answer polynomial $A^s(\alpha)$ in \eqref{answer:polynomial}, i.e.,
\begin{IEEEeqnarray}{l}
\Lambda^s(\alpha)=\sum\limits_{f_1\in[F_1],\ldots,f_M\in[F_M]}\sum\limits_{j\in[\lambda]}\phi_j^{s}(\alpha)\quad\quad\quad\notag\\
\quad\quad\quad\quad\times\bigg(\prod\limits_{m\in[M]}Q_{j}^{(f_m),m,s}(\alpha)\bigg)\cdot\varphi_j^{(f_1,\ldots,f_M)}(\alpha).\IEEEeqnarraynumspace\label{answer:polynomial2}
\end{IEEEeqnarray}

Then, for any $\mathcal{B}\subseteq[N],\mathcal{U}\subseteq[N],|\mathcal{B}|\leq B,|\mathcal{U}|\leq U,\mathcal{B}\cap\mathcal{U}=\emptyset$ where $s\in[S]$, we have
\begin{IEEEeqnarray}{rCl}
0&\leq& I\big(A_{[N]\backslash\mathcal{U}},\theta_m,\mathcal{Z}_m;\mathcal{W},\notag\\
&&\quad\quad\quad\quad\quad\quad\quad\quad\quad\{\theta_{\overline{m}}\}_{\overline{m}\in[M]\backslash\{m\}}|\mathbf{W}^{(\theta_1,\ldots,\theta_M)}\big)\notag\\
&\overset{(a)}{=}&I\big(\{A_{[N]\backslash\mathcal{U}}^{s}\}_{s\in[S]},\theta_m,\mathcal{Z}_m;\mathcal{W},\notag\\
&&\quad\quad\quad\quad\quad\quad\quad\quad\quad\{\theta_{\overline{m}}\}_{\overline{m}\in[M]\backslash\{m\}}|\mathbf{W}^{(\theta_1,\ldots,\theta_M)}\big) \notag\\
&=&I\big(\theta_m,\mathcal{Z}_m;\mathcal{W},\{\theta_{\overline{m}}\}_{\overline{m}\in[M]\backslash\{m\}}|\mathbf{W}^{(\theta_1,\ldots,\theta_M)}\big)\notag\\
&&\quad\quad\quad+I\big(\{A_{[N]\backslash\mathcal{U}}^{s}\}_{s\in[S]};\mathcal{W},\notag\\
&&\quad\quad\quad\quad\quad\;\;\{\theta_{\overline{m}}\}_{\overline{m}\in[M]\backslash\{m\}}|\theta_m,\mathcal{Z}_m,\mathbf{W}^{(\theta_1,\ldots,\theta_M)}\big)\notag\\
&\overset{(b)}{=}&I\big(\{A_{[N]\backslash\mathcal{U}}^{s}\}_{s\in[S]};\mathcal{W},\notag\\
&&\quad\quad\quad\quad\quad\;\{\theta_{\overline{m}}\}_{\overline{m}\in[M]\backslash\{m\}}|\theta_m,\mathcal{Z}_m,\mathbf{W}^{(\theta_1,\ldots,\theta_M)}\big)\notag\\
&\leq&I\big(\{A_{[N]\backslash\mathcal{U}}^{s},A^{s}(\alpha_1),\ldots,A^{s}(\alpha_N)\}_{s\in[S]};\mathcal{W},\notag\\
&&\quad\quad\quad\quad\quad\{\theta_{\overline{m}}\}_{\overline{m}\in[M]\backslash\{m\}}|\theta_m,\mathcal{Z}_m,\mathbf{W}^{(\theta_1,\ldots,\theta_M)}\big)\notag\\
&=&I\big(\{A^{s}(\alpha_1),\ldots,A^{s}(\alpha_N)\}_{s\in[S]};\mathcal{W},\notag\\
&&\quad\quad\quad\quad\quad\{\theta_{\overline{m}}\}_{\overline{m}\in[M]\backslash\{m\}}|\theta_m,\mathcal{Z}_m,\mathbf{W}^{(\theta_1,\ldots,\theta_M)}\big)\notag\\
&&+I\big(\{A_{[N]\backslash\mathcal{U}}^{s}\}_{s\in[S]};\mathcal{W},\{\theta_{\overline{m}}\}_{\overline{m}\in[M]\backslash\{m\}}|\theta_m,\mathcal{Z}_m,\notag\\
&&\quad\quad\quad\quad \{A^{s}(\alpha_1),\ldots,A^{s}(\alpha_N)\}_{s\in[S]},\mathbf{W}^{(\theta_1,\ldots,\theta_M)}\big)\notag\\
&\overset{(c)}{=}&I\big(\{A^{s}(\alpha_1),\ldots,A^{s}(\alpha_N)\}_{s\in[S]};\mathcal{W},\notag\\
&&\quad\quad\quad\quad\quad\{\theta_{\overline{m}}\}_{\overline{m}\in[M]\backslash\{m\}}|\theta_m,\mathcal{Z}_m,\mathbf{W}^{(\theta_1,\ldots,\theta_M)}\big)\notag\\
&&+I\big(\{A_{\mathcal{B}}^{s}\}_{s\in[S]};\mathcal{W},\{\theta_{\overline{m}}\}_{\overline{m}\in[M]\backslash\{m\}}|\theta_m,\mathcal{Z}_m,\notag\\
&&\quad\quad\quad\quad\{A^{s}(\alpha_1),\ldots,A^{s}(\alpha_N)\}_{s\in[S]},\mathbf{W}^{(\theta_1,\ldots,\theta_M)}\big)\notag\\
&\overset{(d)}{=}&I\big(\{A^{s}(\alpha_1),\ldots,A^{s}(\alpha_N)\}_{s\in[S]};\mathcal{W},\notag\\
&&\quad\quad\quad\quad\quad\{\theta_{\overline{m}}\}_{\overline{m}\in[M]\backslash\{m\}}|\theta_m,\mathcal{Z}_m,\mathbf{W}^{(\theta_1,\ldots,\theta_M)}\big)\notag\\
&\overset{(e)}{=}&I\big(\big\{A^{s}(\alpha):\alpha\in\{\beta_{i,s}\}_{i\in[\lambda]}\notag\\
&&\quad\quad\quad\cup\{\alpha_i\}_{i\in[K+X+T_1+\ldots+T_M-1]}\big\}_{s\in[S]};\mathcal{W},\notag\\
&&\quad\quad\quad\quad\quad\{\theta_{\overline{m}}\}_{\overline{m}\in[M]\backslash\{m\}}|\theta_m,\mathcal{Z}_m,\mathbf{W}^{(\theta_1,\ldots,\theta_M)}\big)\notag\\
&\overset{(f)}{=}&I\big(\{w_{i,s}^{(\theta_1,\ldots,\theta_M)}\}_{i\in[\lambda],s\in[S]},\notag\\
&&\quad\quad\{\Lambda^{s}(\alpha_i)+z_i^{s}\}_{i\in[K+X+T_1+\ldots+T_M-1],s\in[S]};\mathcal{W},\notag\\
&&\quad\quad\quad\quad\quad\{\theta_{\overline{m}}\}_{\overline{m}\in[M]\backslash\{m\}}|\theta_m,\mathcal{Z}_m,\mathbf{W}^{(\theta_1,\ldots,\theta_M)}\big)\notag\\
&\overset{(g)}{=}&I\big(\{\Lambda^{s}(\alpha_i)+z_i^{s}\}_{i\in[K+X+T_1+\ldots+T_M-1],s\in[S]};\mathcal{W},\notag\\
&&\quad\quad\quad\quad\{\theta_{\overline{m}}\}_{\overline{m}\in[M]\backslash\{m\}}|\theta_m,\mathcal{Z}_m,\mathbf{W}^{(\theta_1,\ldots,\theta_M)}\big)\label{terms:1}\\
&\overset{(h)}{=}&0, \notag
\end{IEEEeqnarray}
where $(a)$ follows by \eqref{answers:gene};
$(b)$ holds because $\theta_m,\mathcal{Z}_m$ are generated independently of $\mathcal{W},\{\theta_{\overline{m}}\}_{\overline{m}\in[M]\backslash\{m\}}$ by \eqref{reconstruction} and \eqref{independence:entry};
$(c)$ is due to the fact that the answer $A_{n}^{s}$ is equivalent to $A^{s}(\alpha_n)$ for any authentic server $n\in[N]\backslash(\mathcal{U}\cup\mathcal{B})$ by \eqref{answers}-\eqref{answer:polynomial};
$(d)$ follows by the fact that the Byzantine servers maliciously return arbitrary responses $\{A_{\mathcal{B}}^{s}\}_{s\in[S]}$ from $\mathbb{F}_q$ to confuse the users and thus the answers of Byzantine servers cannot leak anything to the users \cite{Tajeddine222,Wang123,wang-byzantine};
$(e)$ holds because $A^{s}(\alpha)$ is a polynomial of degree $\lambda+K+X+T_1+\ldots+T_M-2$ such that $\{A^{s}(\alpha_1),\ldots,A^{s}(\alpha_N)\}$ and $\{A^{s}(\alpha):\alpha\in\{\beta_{i,s}\}_{i\in[\lambda]}\cup\{\alpha_i\}_{i\in[K+X+T_1+\ldots+T_M-1]}\}$ are determined of each other by Lagrange interpolation rules and P2-P4 for any $s\in[S]$;
$(f)$ follows by \eqref{evaluating:1}, \eqref{answer:polynomial}, \eqref{symmetric:122} and \eqref{answer:polynomial2};
$(g)$ is due to \eqref{file symbols} and \eqref{chose:parameters};
$(h)$ follows from the fact that $\{z_i^{s}\}_{i\in[K+X+T_1+\ldots+T_M-1],s\in[S]}$ are i.i.d. uniformly over $\mathbb{F}_q$ and are generated independently of all other variables in \eqref{terms:1}.

This verified the server privacy and blind privacy \eqref{Infor:priva cons2}.
\end{IEEEproof}


\end{appendix}


\end{document}